\documentclass[twocolumn]{aastex62}  

\newcommand\apjcls{1}
\newcommand\aastexcls{2}
\newcommand\othercls{3}

\newcommand\papercls{\aastexcls}

\if\papercls \apjcls
\usepackage{apjfonts}
\else\if\papercls \othercls
\usepackage{epsfig}
\usepackage{margin}
\usepackage{times}
\fi\fi
\usepackage{ifthen}
\usepackage{natbib}
\usepackage{amssymb, amsmath}
\usepackage{appendix}
\usepackage{etoolbox}
\usepackage[T1]{fontenc}
\usepackage{paralist}

\if\papercls \apjcls
\newcommand\aas{\ref@jnl{AAS Meeting Abstracts}}
\newcommand\dps{\ref@jnl{AAS/DPS Meeting Abstracts}}
\newcommand\maps{\ref@jnl{MAPS}}
\else\if\papercls \othercls
\usepackage{astjnlabbrev-jh}
\fi\fi

\if\papercls \aastexcls
\hypersetup{citecolor=blue, 
            linkcolor=blue, 
            menucolor=blue, 
            urlcolor=blue}  
\else
\usepackage[
bookmarks=true,           
bookmarksnumbered=true,   
colorlinks=true,          
citecolor=blue,           
linkcolor=blue,           
menucolor=blue,           
urlcolor=blue,            
linkbordercolor={0 0 1},  
pdfborder={0 0 1},
frenchlinks=true]{hyperref}
\fi

\if\papercls \othercls

\else

\fi

\providecommand{\adsurl}[1]{\href{#1}{ADS}}

\makeatletter
\patchcmd{\NAT@citex}
  {\@citea\NAT@hyper@{%
     \NAT@nmfmt{\NAT@nm}%
     \hyper@natlinkbreak{\NAT@aysep\NAT@spacechar}{\@citeb\@extra@b@citeb}%
     \NAT@date}}
  {\@citea\NAT@nmfmt{\NAT@nm}%
   \NAT@aysep\NAT@spacechar\NAT@hyper@{\NAT@date}}{}{}

\patchcmd{\NAT@citex}
  {\@citea\NAT@hyper@{%
     \NAT@nmfmt{\NAT@nm}%
     \hyper@natlinkbreak{\NAT@spacechar\NAT@@open\if*#1*\else#1\NAT@spacechar\fi}%
       {\@citeb\@extra@b@citeb}%
     \NAT@date}}
  {\@citea\NAT@nmfmt{\NAT@nm}%
   \NAT@spacechar\NAT@@open\if*#1*\else#1\NAT@spacechar\fi\NAT@hyper@{\NAT@date}}
  {}{}
\makeatother

\makeatletter
\DeclareRobustCommand{\lowcase}[1]{\@lowcase#1\@nil}
\def\@lowcase#1\@nil{\if\relax#1\relax\else\MakeLowercase{#1}\fi}
\makeatother

\DeclareSymbolFont{UPM}{U}{eur}{m}{n}
\DeclareMathSymbol{\umu}{0}{UPM}{"16}
\let\oldumu=\umu
\renewcommand\umu{\ifmmode\oldumu\else\math{\oldumu}\fi}

\if\papercls \othercls

\else

\fi

\let\oldsim=\sim
\renewcommand\sim{\ifmmode\oldsim\else\math{\oldsim}\fi}
\let\oldpm=\pm
\renewcommand\pm{\ifmmode\oldpm\else\math{\oldpm}\fi}
\newcommand\by{\ifmmode\times\else\math{\times}\fi}

\newbox{\wdbox}
\renewcommand\c{\setbox\wdbox=\hbox{,}\hspace{\wd\wdbox}}
\renewcommand\i{\setbox\wdbox=\hbox{i}\hspace{\wd\wdbox}}

\newcount\timect
\newcount\hourct
\newcount\minct
\newcommand\now{\timect=\time \divide\timect by 60
         \hourct=\timect \multiply\hourct by 60
         \minct=\time \advance\minct by -\hourct
         \number\timect:\ifnum \minct < 10 0\fi\number\minct}

\catcode`@=11

\newcommand\comment[1]{}

\newcommand\commenton{\catcode`\%=14}

\renewcommand\math[1]{$#1$}
\newcommand\mathshifton{\catcode`\$=3}

\let\atab=&
\newcommand\atabon{\catcode`\&=4}

\let\oldmsp=\sp
\let\oldmsb=\sb
\def\sp#1{\ifmmode
           \oldmsp{#1}%
         \else\strut\raise.85ex\hbox{\scriptsize #1}\fi}
\def\sb#1{\ifmmode
           \oldmsb{#1}%
         \else\strut\raise-.54ex\hbox{\scriptsize #1}\fi}
\newbox\@sp
\newbox\@sb
\def\sbp#1#2{\ifmmode%
           \oldmsb{#1}\oldmsp{#2}%
         \else
           \setbox\@sb=\hbox{\sb{#1}}%
           \setbox\@sp=\hbox{\sp{#2}}%
           \rlap{\copy\@sb}\copy\@sp
           \ifdim \wd\@sb >\wd\@sp
             \hskip -\wd\@sp \hskip \wd\@sb
           \fi
        \fi}
\def\msp#1{\ifmmode
           \oldmsp{#1}
         \else \math{\oldmsp{#1}}\fi}
\def\msb#1{\ifmmode
           \oldmsb{#1}
         \else \math{\oldmsb{#1}}\fi}

\def\supon{\catcode`\^=7}

\def\subon{\catcode`\_=8}

\def\supsubon{\supon \subon}

\newcommand\actcharon{\catcode`\~=13}

\newcommand\paramon{\catcode`\#=6}

\comment{And now to turn us totally on and off...}

\newcommand\reservedcharson{ \commenton  \mathshifton  \atabon  \supsubon 
                             \actcharon  \paramon}

\catcode`@=12
\reservedcharson

\if\papercls \apjcls

\else

\fi

\newcommand\chisq{\ifmmode{\chi\sp{2}}\else\math{\chi\sp{2}}\fi}
\newcommand\redchisq{\ifmmode{ \chi\sp{2}\sb{\rm red}}
                    \else\math{\chi\sp{2}\sb{\rm red}}\fi}
\newcommand\Teq{\ifmmode{T\sb{\rm eq}}\else$T$\sb{eq}\fi}
\newcommand\Teff{\ifmmode{T\sb{\rm eff}}\else$T$\sb{eff}\fi}
\newcommand\mjup{\ifmmode{M\sb{\rm Jup}}\else$M$\sb{Jup}\fi}
\newcommand\rjup{\ifmmode{R\sb{\rm Jup}}\else$R$\sb{Jup}\fi}
\newcommand\msun{\ifmmode{M\sb{\odot}}\else$M\sb{\odot}$\fi}
\newcommand\rsun{\ifmmode{R\sb{\odot}}\else$R\sb{\odot}$\fi}
\newcommand\mearth{\ifmmode{M\sb{\oplus}}\else$M\sb{\oplus}$\fi}
\newcommand\rearth{\ifmmode{R\sb{\oplus}}\else$R\sb{\oplus}$\fi}

\usepackage{amsmath}
\usepackage{graphicx}

\usepackage{epsfig}
\usepackage{natbib}
\usepackage{wasysym}
\usepackage{lipsum}
\usepackage{mathrsfs}
\usepackage{float}
\usepackage{rotating}
\usepackage{graphicx}
\usepackage{epstopdf}
\usepackage{colortbl}
\providecommand{\e}[1]{\ensuremath{\times 10^{#1}}}

\usepackage{color}

\definecolor{twitterblue}{RGB}{64,153,255}

\usepackage{hyperref}

\usepackage{lineno}

\usepackage{filecontents}

\begin{document}

\shorttitle{Stellar Flares \& Exo-Earth's Climate}
\shortauthors{Chen, H., De Luca, P., Hochman, A., \& Komacek, T.D.}

\title{{\bf Effects of transient stellar emissions on planetary climates of tidally-locked exo-earths}}

\author[0000-0003-1995-1351]{Howard Chen}
\affil{Department of Aerospace, Physics, and Space Sciences, Florida Institute of Technology, Melbourne, FL 32901}

\affil{Sellers Exoplanet Environments Collaboration (SEEC), NASA Goddard Space Flight Center, Greenbelt, MD 20771}

\author[0000-0002-0416-4622]{Paolo De Luca}
\affil{Barcelona Supercomputing Center (BSC), Barcelona, Spain}

\author[0000-0002-9881-1893]{Assaf Hochman}

\affil{Fredy and Nadine Herrmann Institute of Earth Sciences, Hebrew University of Jerusalem, Jerusalem, Israel}

\author[0000-0002-9258-5311]{Thaddeus D. Komacek}
\affiliation{Department of Physics (Atmospheric, Oceanic and Planetary Physics), University of Oxford, Oxford OX1 3PU, UK}
\affiliation{Blue Marble Space Institute of Science, Seattle, WA 98104, USA}
\affiliation{Department of Astronomy, University of Maryland, College Park, MD 20742, USA}

\correspondingauthor{Howard Chen} \email{hchen@fit.edu}

\begin{abstract}
Space weather events in planetary environments sourced from transient host star emissions, including stellar flares, coronal mass ejections, and stellar proton events, can substantially influence an exoplanet’s climate and atmospheric evolution history. These time-dependent events may also affect our ability to measure and interpret its properties by modulating reservoirs of key chemical compounds and changing the atmosphere's brightness temperature. The majority of previous work focusing on photochemical effects, ground-level UV dosages, and consequences on observed spectra. Here, using three-dimensional (3D) general circulation models with interactive photochemistry, we simulate the climate and chemical impacts of stellar energetic particle events and periodic enhancements of UV photons. We use statistical methods to examine their effects on synchronously rotating TRAPPIST-1e-like planets on a range of spatiotemporal scales. We find that abrupt thermospheric cooling is associated with radiative cooling of NO and CO$_2$, and middle-to-lower atmospheric warming is associated with elevated infrared absorbers such as N$_2$O and H$_2$O. In certain regimes, in particular for climates around moderately active stars, atmospheric temperature changes are strongly affected by O$_3$ variability. Cumulative effects are largely determined by the flare frequency and the instantaneous effects are dependent on the flare’s spectral shape and energy.   In addition to effects on planetary climate and atmospheric chemistry, we find that intense flares can energize the middle atmosphere, causing enhancements in wind velocities up to 40 m s$^{-1}$ in sub-stellar night-sides between 30 and 50 km in altitude. Our results suggest that successive, more energetic eruptive events from younger stars may be a pivotal factor in determining the atmosphere dynamics of their planets.
\end{abstract}

\section{Introduction} 
\label{sec:intro}
Space weather events originating from a star, including X-ray and extreme UV radiation, coronal mass ejections (CME) and stellar energetic particle (SEP) precipitation, will likely impede or disturb the atmospheres of rocky exoplanets (see, e.g., \citealt{DongEt2017ApJL,france2020}). Analyses of the Kepler data show that the average flare energy detected is ${\sim}10^{35}$ erg \citep{DavenportEt2016ApJ}, and the average flare amplitude does not change significantly between G-K-M stars. In contrast, flare frequency and the spectral distributuon of flares greatly depends on the host type \citep{van2017}, and the latter is found to vary even within the same spectral class \citep{paudel2024}. From the Transiting Exoplanet Satellite Survey (TESS) catalog, it has been shown that 97\% of the most intense flares released energies of $>10^{33}$ erg \citep{howard2022,vida2021,stelzer2022}, with a typical TESS star having daily flares with optical energies between $10^{32}–10^{33}$ erg \citep{ealy2024}.  Meanwhile, the level of magnetic activity in terms of the energy and frequency of super-flare events with energies between $10^{33}$ and $10^{35}$ erg drops down within 600 Myrs for G-type stars and 1 Gyr for K-dwarfs and M-dwarfs \citep{davenport2019,ilin2021}. The latter hosts include the well-characterized TRAPPIST-1 system, which, due to their deep convective zones and longer spin-down timescales, are known to retain a high level of magnetic and chromospheric activity for Gyrs (e.g., \citealt{baraffe2015}). This can affect atmospheric environments of close-in exoplanets on long timescales, inducing water loss via photolysis and hydrogen escape (e.g., \citealt{louca2023,fromont2024}). However, the detailed flare characteristics of the youngest of the samples ($< 300$ Myr) are still being investigated (see, e.g., \citealt{feinstein2024,howard2024}).

Stellar eruptive events have been shown to dramatically influence the abundance of major and minor chemical species in the atmospheres of habitable zone (HZ) planets, particularly for those around M-dwarfs. The most significant effects on Earth-like, or strongly oxygenated, planets are those on the ozone column abundances \citep{chen2021,ridgway2023}. For a single large AD Leonis-like flare, the impact of ozone erosion is minimal as the ozone layer is recovered over just a few Earth years \citep{SeguraEt2010AsBio}. However, the ozone layer is rapidly eroded even with moderate flare energy and proton fluence assumptions \citep{TilleyEt2019AsBio} for repeated events with frequencies typical for Kepler stars.   The effects on other photochemical constituents, for example, nitric acid or HNO$_3$ (e.g., \citealt{HerbstEt2019A&Aa}), depending on the assumed energetics of the associated primary protons precipitated in the atmosphere of the planet, typically between 16 MeV and 0.5 TeV \citep{scheucher2020}.

Others, using a combination of observational reconstruction and numerical modeling, have found that secular injection of energetic particles can lead to enhancements of greenhouse gases such as water vapor and nitrogen oxides, i.e., NO$_2$ and N$_2$O, during large stellar proton events \citep{airapetian2016,chen2021}. Such enhancement may have important implications for surface climate, for example, by causing abrupt surface warming that could alter previous conclusions regarding surface habitability. As implied by other 1-D models, sudden and periodic enhancements in the abundance of other greenhouse gasses or UV absorbers by stellar flares can be substantial (up to 4-5 orders of magnitude change in their mixing ratios), but further efforts to quantify these processes using modeling tools at the global (3-D) level are needed. In CO$_2$-rich conditions, an increase in upper atmospheric CO$_2$ caused by repeated flaring in oxidizing atmospheres can dramatically sculpt the temperatures of upper atmospheres around M-dwarfs through CO$_2$ radiative cooling \citep{WordsworthEt2014ApJL}. In these environments, repeated flaring may initiate an entirely different chemical reaction chain that may have substantial implications for surface climate. Ammonia, for example, which has been seen as a potential greenhouse gas to warm the surfaces of early Earth and potentially Earth-like planets \citep{sagan1972}, would have severely limited abundances due to the absence of strong UV absorbers between 200 and 300 nm that would prevent photo-destruction of ammonia in the atmosphere. 

These 1-D investigations show, on a theoretical level\footnote{Stellar activity-driven photochemical effects have been observed by the JWST (e.g., \citealt{tsai2023,rustamkulov2023}), but such processes may be challenging to directly measure by near-term instruments on habitable zone planets with $R_p \lesssim 2~R_\oplus$.}, that photochemical effects as a result of time-dependent stellar flares cannot be neglected in efforts to more realistically characterize the composition of astrobiological significant trace gases in rocky exoplanet atmospheres. It is, therefore, essential to further examine their impacts on habitability and exoplanet temporal variability and climate.  Planetary climate dynamics can be an important factor in determining the observable properties of exoplanets because they set the three-dimensional atmospheric temperature structure, mixing of chemical species and aerosols, and resulting cloud distributions. An understanding of planetary climate dynamics is needed to interpret spectroscopic observations of exoplanet atmospheres. The precise characteristics of an exoplanet's atmosphere  are linked to its orbital and rotation period around an M-dwarf star. Due to the dependence of the rotation period on the stellar spectral type, different rotational states lead to diverse climate evolution pathways of their surface temperatures (e.g., \citealt{merlis2010,yang2013stabilizing,CaroneEt2018MNRAS,turbet2016,kumar2016inner,kumar2017,WayEt2016GRL,komacek2019,way2020,he2022,chen2023}), as well as divergent chemical distributions (e.g., \citealt{luo2023,cohen2022,braam2025}).

There has yet to be a systematic study of the effect of stellar flaring events on the planetary climate and atmospheric dynamics of planets orbiting M-dwarf stars, as the majority of previous work has focused on understanding effects on stratospheric chemistry using 1D photochemistry models. Exoplanet climate forcings due to external agents such as astrophysical events, a relatively unexplored area of research, could have major ramifications for both habitability and observations. Studies on the  relationship between Earth and the Sun during strong geomagnetic storms found a ${\sim}$0.5 K cooling in the middle stratosphere over the tropics and up to 2 K over southern high latitudes \citep{rozanov2005}. The most recent investigations into the effects of extreme solar energetic events underscored the key role of the geomagnetic field strength in moderating instantaneous and long-term effects on the atmosphere \citep{arsenovic2024}.  Specifically,  during historical periods of weakened planetary magnetic fields,  their global climate models have shown that  odd nitrogen species can remain in the atmosphere on the order of several years in the lower latitudes,  leading to ozone losses on a decadal timescale and hence markedly increased ground-level UV radiation.

Another study on the effects of the AD 775 great solar event on Earth’s climate has shown ozone depletion followed by a decrease in near-surface air temperature by up to 4 K \citep{sukhodolov2017}. This temperature anomaly then affects the Polar-Night Jet Oscillation and accelerates the tropospheric mean flow soon after the event in December. Their model also found that a more pronounced meridional circulation leads to warming in the Earth's polar regions and the northern hemisphere. The most recent investigations into the effects of extreme solar energetic events on Earth's particle and atmospheric environment underscored the key role of the geomagnetic field strength in determining instantaneous and long-term effects on the atmosphere \citep{arsenovic2024}, echoing earlier results on strongly to weakly magnetized exoplanets \citep{chen2021}. These terrestrial studies indicate that climate effects may also be important for extrasolar environments, in particular for activities from cool stars extending up to higher energy domains compared to solar measurements.

Here, we simulate the putative climate dynamics of TRAPPIST-1e analogs, assuming Earth-like atmospheres with different levels of stellar flaring activity derived from observational measurements of M-dwarfs. We analyze the flare-driven variability in the thermal structure and dynamics of the atmosphere and their correlation with the variability in the abundance and distribution of key chemical constituents. This manuscript is organized as follows: We describe our GCM modeling framework and statistical analyses in Section 2. We present results statistically quantifying the impact of varying levels of flaring activity on climate and atmospheric chemistry in Section 3. Finally, Section 4 discusses our findings and places this work in the broader context of climate and atmospheric evolution models.

\section{Methods \& Numerical Model}
\subsection{3D Global Chemistry-Climate System Modeling}
\label{sec:method}
We use the National Center for Atmospheric Research's Community Earth System Model (CESM) version 1.2.2 \citep{neale2010description,MarshEt2013JGR} to simulate the atmospheres of template exo-Earths under the influence of stellar eruptive events, including both the effects of stellar UV flares and stellar energetic particle precipitation. The Whole Atmosphere Community Climate Model (WACCM), a supercomponent set of CESM, is a 3-D global climate model that simulates atmospheric chemistry, radiation, thermodynamics, and dynamics. In particular, the model includes chemical speciation, network reaction, and transport processes necessary to fully account for the effects of time-dependent stellar events that are of interest here. We follow previous uses of WACCM in exoplanet studies (e.g., \citealt{ChenEt2019ApJ,afentakis2023,luo2023}) and specify the Community Atmosphere Model v4 (CAM4) as the atmosphere component of WACCM. CAM4 employs the native Community Atmospheric Model Radiative Transfer (CAMRT) radiation scheme \citep{kiehl1983co2}, the Hack scheme for shallow convection \citep{Hack1994JGR}, the Zhang–McFarlane scheme for deep convection \citep{zhang1995sensitivity}, and the Rasch–Kristjansson scheme for condensation, evaporation, and precipitation \citep{zhang2003modified}. Note that the specific atmospheric component and the convection parameterization scheme have been shown to substantially influence the simulated climates of planets around M-dwarfs \citep{bin2018,sergeev2020}.

We conduct four independent simulations with WACCM initialized from the same steady state solution. These simulations have   TRAPPIST-1e's planetary mass and radius (E. Agol et al. 2021), Earth-similar compositions, Earth’s present-day insolation, but with slightly different transient stellar emissions, that is, time-varying UV spectra and input proton fluence.  We also assumed modern Earth's surface albedo, surface topography, and landmass distribution to facilitate comparison with previous work studying ocean-covered (exo)planets. Previous exoplanet applications of WACCM simulations have also assumed Earth's surface properties in part due to the numerical challenges of replacing ocean bathymetries of aqua-planets with the existing model land masks (see, e.g., \citealt{luo2023,ChenEt2019ApJ}). 

We use an open-source Python flare toolkit developed by \citet{LoydEt2018ApJa}  to compute flare spectra and flare light curves. This stochastic flare model is based on the MUSCLES Treasury Survey V, which we use to generate UV light curves. The MUSCLES survey (HST observing program 13650) characterized the low-mass stellar radiation environment, including X-ray, XUV, FUV, and NUV fluxes \citep{FranceEt2016ApJ}. Model flares are stochastically generated, depending on the range of user-specified energies and frequencies, from observed data drawn from two stellar populations: the MUSCLE M-dwarf sample and four active stars Proxima Centauri, AD Leo, EV Lac, and AU Mic. The UV flare spectra are obtained by multiplying the quiescent UV spectra of each stellar SED with the active-to-quiescent flux ratio given by the flare generator. However, we note that energetic particles, or stellar proton events, are found to be responsible for $\sim 99$\% of the photochemical effects \citep{SeguraEt2010AsBio,TilleyEt2019AsBio,GrenfellEt2012AsBio}. As the estimated color temperatures of M-dwarf flares range between 7700 and 14000 K \citep{KowalskiEt2013ApJS}, here all of our modeled flaring events assume black-body temperatures of 9000 K. However, M-dwarf flares at much hotter temperatures have been observed (e.g., \citealt{osten2016}). 

To compute the associated energetic particle precipitation in the ambient atmosphere, we follow previous studies using solar scaling of an event’s proton fluence based on near-Earth satellite data, assuming a linear relationship
between the proton flux and production rate of ion pairs that should be valid for the most Earth-like atmospheres \citep{jackman2008short}. We assume that all of the particles are protons. We compute the expected peak proton fluences from the SiIV energy of stellar flares using the prescription of \citet{YoungbloodEt2017ApJ}. The proton energy spectra for each event are assumed to be fixed and consistent with the 2003 Halloween solar proton event \citep{Funke:2011}. Although inappropriate for the most extreme events and M-dwarf activity \citep{hu2022,HerbstEt2019A&Aa}, this assumption should not deviate from the typical events expected from our chosen flare energy range. Across the four numerical experiments conducted, we assume three different levels of flaring activity of TRAPPIST-1. We summarize these inputs in Table~\ref{table1}. The template ‘quiescent’ simulation uses a TRAPPIST-1 spectra with its UV regime ($\lambda < 125$ nm) replaced with an inactive PHOENIX BT-Settl UV model. The 'Moderate' simulation uses the out-of-the-box flare frequency for sample MUSCLES stars ($\alpha  = 0.76$; \citealt{LoydEt2018ApJa,FranceEt2016ApJ}). The 'Extreme' simulations assume a cumulative flare index $\alpha  = 0.58$, indicative of more magnetically active stars \citep{LoydEt2018ApJa,FranceEt2016ApJ}. Both these simulations have flare rates $\mu = 6$ day$^{-1}$.  Lastly, the `Active' used a power-law index of $\alpha  = 0.52$  and a reduced flare rate of $\mu = 4$ day$^{-1}$. This choice in flare statistic is in closer agreement with the sample of actively flaring stars (Table 6 in \citealt{LoydEt2018ApJa}). The latter simulation set aims to test the effect of a slightly higher frequency in sampling more energetic flares, but a lower flare rate {\it overall} across all energies. For the three non-quiescent simulation sets, we restrict their input flare energies to between $10^{30}$ and $10^{34.5}$ erg, and all model data are outputted at daily cadences equal to one Earth day. Table 1 summarizes the stellar and planetary parameters in this study.﻿

\citet{jackman2023} showed that the assumed blackbody flare temperature value (i.e., 9000 K\footnote{(This temperature value is also a common assumption in previous observational and theory work (see e.g., \citealt{kowalski2024time,howard2020}) and have been used as lower limits in the best-model temperature models \citep{loyd2018}.}) underestimates the flare energies of field-age M dwarfs by a factor of 6.5 in the NUV and 30 in the FUV range. Further, \citet{kowalski2024} found that, in many cases, single blackbodies does a poor job in representing the NUV flare spectra. Instead, the continuum flux  increase toward shorter wavelengths. Therefore caution must be placed when interpreting our numerical results, as higher flare temperatures would impact photodissociation rates and enhance/mute the impacts of stellar protons. For example, the NUV regime is largely responsible for ozone photolysis, whereas the FUV contains key bands for ozone production. The precise spectral shape of each flare determines the production and loss of cooling/warming agents such as NO and N$_2$O. How different flare temperatures might change spectral windows and how UV irradiation intersects with the effects of ionizing particles remains to be seen. Exploring alternative  temperatures and an expanded parameter space covering flare frequency distributions of earlier type stars, is an important next step for both 1-D photochemistry and 3-D climate models.

\subsection{Statistical Analyses}

We test the statistical significance of the field medians by applying a two-tailed Wilcoxon rank-sum test \citep{mann1947}. The test does not assume data to be normally distributed, and it checks the null hypothesis that the field median of the Extreme and Moderate simulation is equal to the field median of the Quiescent simulation at the 1\% significance level (p-value < 0.01). 

For the above four sets of climate model simulations, we calculate boxplots of the anomalies for regions as follows: equatorial substellar (-15$^\circ$-15$^\circ$N, 165$^\circ$-195$^\circ$E), equatorial antistellar (-15$^\circ$-15$^\circ$N, 345$^\circ$-15$^\circ$E), midlatitude substellar (30$^\circ$-60$^\circ$N, 165$^\circ$-195$^\circ$E), midlatitude antistellar (30$^\circ$-60$^\circ$N, 345$^\circ$-15$^\circ$E).  We select these subregions because they divide the day-side and night-side hemispheres, including the terminator regions, better to identify the effects of tidal locking on flare activity. The anomalies were computed concerning the Quiescent simulation. Specifically, we first take the field medians of the Extreme, Moderate, and Active simulations and then subtract from them the field and temporal medians of the Quiescent simulation. In this case, we apply a Kruskal-Wallis H test (or one-way ANOVA on ranks; \citealt{kruskal1952}), which is an extension of the Wilcoxon rank-sum test to three groups, to check whether at least one population anomaly median of one group is different from the population median of at least one other group at the 1\% significance level.

Along with field medians and boxplots, we also compute monthly 8-year median time-series for all four simulated cases over the entire planet, day-side (-90$^\circ$-90$^\circ$N, 90$^\circ$-270$^\circ$E) and night-side (-90$^\circ$-90$^\circ$N, 270$^\circ$-90$^\circ$E). Lastly, we calculate the vertical difference profile of wind speed (m s$^{-1}$) over the region -15$^\circ$-15$^\circ$N, 0$^\circ$-50$^\circ$E. This region was chosen as it showed one of the largest anomalies in the field median analyses. As for the other cases, we apply a two-tailed Wilcoxon rank-sum test to infer whether the temporal and field medians are statistically different at the 1\% significance level.

\begin{table*}[t]
\centering
\begin{tabular}{ |c||c| }
    \hline
    \textbf{Stellar or Planetary Parameter} & \textbf{Value/Range} (units) \\
    \hline
    Planetary Orbital Period  & 4.62 (d) \\
    Stellar $T_{\rm eff}$ & 2600 (K) \\
    Incident Flux & 900 (W m$^{-2}$)\\
    Planet Radius & 5733 (km) \\
    Proton Energy Spectra & 2003 October-November SPE$^1$ \\
    Cumulative Flare Index & 0.52, 0.58, 0.76$^2$ \\
    Flare Rate & 3-6 day$^{-1}$ \\
    Flare Energy & $10^{30} - 10^{34.5}$ ergs \\
    Flare Blackbody Temp.  &  9000 K \\
    Blackbody SiIV Energy Ratio & 160 \\
    SiIV Quiescent Flux & 0.1 ergs s$^{-1}$ cm$^{-2}$ \\
    \hline
\end{tabular}
\caption{Ranges of initial stellar and planetary properties explored in the numerical simulations. $^2$The 'Moderate' experiment refers to those with the cumulative flare index set to $\alpha = 0.76$; the Quiescent refers to those with the variability of the stellar spectral energy distribution turned off, and the 'Extreme' is with a lower flare index $\alpha = 0.58$ and increased flare rate. These estimates of their flare frequency and flare indices are based on the HST MUSCLES survey \citep{loyd2018,FranceEt2016ApJ}. $^1$\citet{LopezEt2005JGR,jackman2008short}}
\label{table1}
\end{table*}

\begin{table*}[h!]
\centering
\begin{tabular}{ ||>{\columncolor[gray]{0.8}} c|c| |c|c| }
    \hline
    \textbf{Value} & \textbf{Atm. Level (mbar) } & \textbf{Atm. Variable } & \textbf{Case Name} \\
    \hline
654.88 K & 1.223\e{-5}    & Temperature  & Moderate   \\
204.87 K & 0.008 & Temperature   & Moderate   \\
62.39 m s$^{-1}$ & 1.223\e{-5}    & Wind Speed  & Moderate   \\
28.35 m s$^{-1}$ & 0.008 & Wind Speed  & Moderate   \\
4.10\e{-10}    & 1.223\e{-5}    & OH vmr  & Moderate   \\
1.18\e{-8}    & 0.008 & OH vmr & Moderate   \\
1.83\e{-6}    & 1.223\e{-5}    & H$_2$O vmr & Moderate   \\
1.64\e{-5}    & 0.008 & H$_2$O vmr & Moderate   \\
3.04\e{-7}    & 1.223\e{-5}    & N$_2$O vmr & Moderate   \\
2.55\e{-6}    & 0.008 & N$_2$O vmr & Moderate   \\
1.09\e{-13}    & 1.223\e{-5}     & NO$_2$ vmr & Moderate   \\
1.90\e{-7}    & 0.008 & NO$_2$ vmr & Moderate   \\
2.1\e{-4} & 1.223\e{-5}    & NO vmr  & Moderate   \\
3.18\e{-6}    & 0.008 & NO vmr  & Moderate   \\
5.14\e{-10}    & 1.223\e{-5}     & O$_3$ vmr  & Moderate   \\
2.59\e{-8}    & 0.008 & O$_3$ vmr  & Moderate   \\
761.76 K & 1.223\e{-5}    & Temperature   & Quiescent \\
210.25 K & 0.008 &  Temperature   & Quiescent \\
44.39 m s$^{-1}$ & 1.223\e{-5}   & Wind Speed & Quiescent \\
24.97 m s$^{-1}$ & 0.008 & Wind Speed   & Quiescent \\
2.25\e{-10}    & 1.223\e{-5}     & OH vmr  & Quiescent \\
2.10\e{-9}    & 0.008 & OH vmr  & Quiescent \\
1.95\e{-6}    & 1.223\e{-5}   & H$_2$O vmr& Quiescent \\
6.85\e{-8}    & 0.008 & H$_2$O vmr & Quiescent \\
1.55\e{-9}    & 1.223\e{-5}    & N$_2$O vmr & Quiescent \\
1.81\e{-08} & 0.008   & N$_2$O vmr & Quiescent \\
5.34\e{-14}    & 1.223\e{-5}    & NO$_2$ vmr & Quiescent \\
1.34\e{-10}    & 0.008 & NO$_2$ vmr & Quiescent \\
1.36\e{-4} & 1.223\e{-5}    & NO vmr  & Quiescent \\
1.29\e{-7}    & 0.008 & NO vmr  & Quiescent \\
4.28\e{-10}    & 1.223\e{-5}    & O$_3$ vmr  & Quiescent \\
3.09\e{-7}    & 0.008 & O$_3$ vmr  & Quiescent \\
603.99 K & 1.223\e{-5}   &  Temperature   & Extreme    \\
240.58 K & 0.008 &  Temperature   & Extreme    \\
62.61  m s$^{-1}$ & 1.223\e{-5}    & Wind Speed  & Extreme    \\
39.92 m s$^{-1}$ & 0.008 & Wind Speed  & Extreme    \\
1.43\e{-9}    & 1.223\e{-5}     & OH vmr  & Extreme    \\
1.34\e{-7}    & 0.008 & OH vmr  & Extreme    \\
1.87\e{-6}    &  1.223\e{-5}   & H$_2$O vmr & Extreme    \\
7.9\e{-4} & 0.008 & H$_2$O vmr & Extreme    \\
5.71\e{-6}    & 1.223\e{-5}    & N$_2$O vmr & Extreme    \\
2.03\e{-5}    & 0.008 & N$_2$O vmr & Extreme    \\
2.02\e{-13}    & 1.223\e{-5}    & NO$_2$ vmr & Extreme    \\
4.34\e{-6}    & 0.008 & NO$_2$ vmr & Extreme    \\
3.81\e{-4}  & 1.223\e{-5}  & NO vmr  & Extreme    \\
2.95\e{-5}    & 0.008 & NO vmr  & Extreme    \\
4.44\e{-10}    & 1.223\e{-5}   & O$_3$ vmr  & Extreme    \\
3.77\e{-9}    & 0.008 & O$_3$ vmr  & Extreme   \\
\hline
\end{tabular}
\caption{Summary of Simulated Results for the Quiescent, Moderate, and Extreme Stellar States. Values recorded at pressures $1.223\e{-5}$ mbar correspond to the lower thermosphere, while those recorded  at  0.008 mbar correspond to the mesosphere. On Earth-like planets and Sun-like stars, the latter region is where energetic particle precipitation is expected to be particularly important \citep{jackman2008short}. The labeling of each input flare characteristic is described in Table~\ref{table1}. vmr = volume mixing ratio. \label{table2}}
\end{table*}

\section{Results}

First, we present the temporal medians of our three simulated cases. We then examine regional anomalies between different hemispheres and vertical domains before delving into the temporal evolution of atmospheric temperature and relevant chemical species. We conclude the study by presenting possible regimes of flare-induced dynamic variability in the atmosphere. Table~\ref{table1} shows selected results, including atmospheric temperature, wind speed, and mixing ratios of gas phase constituents (OH, H$_2$O, N$_2$O, NO$_2$, and O$_3$). WeHere we focus on the variables most relevant to climate dynamics in the the main figures, whereas the Appendix provides information on the photochemically important species.

\begin{figure*}[t] 
\begin{center}
\includegraphics[width=1.8\columnwidth]{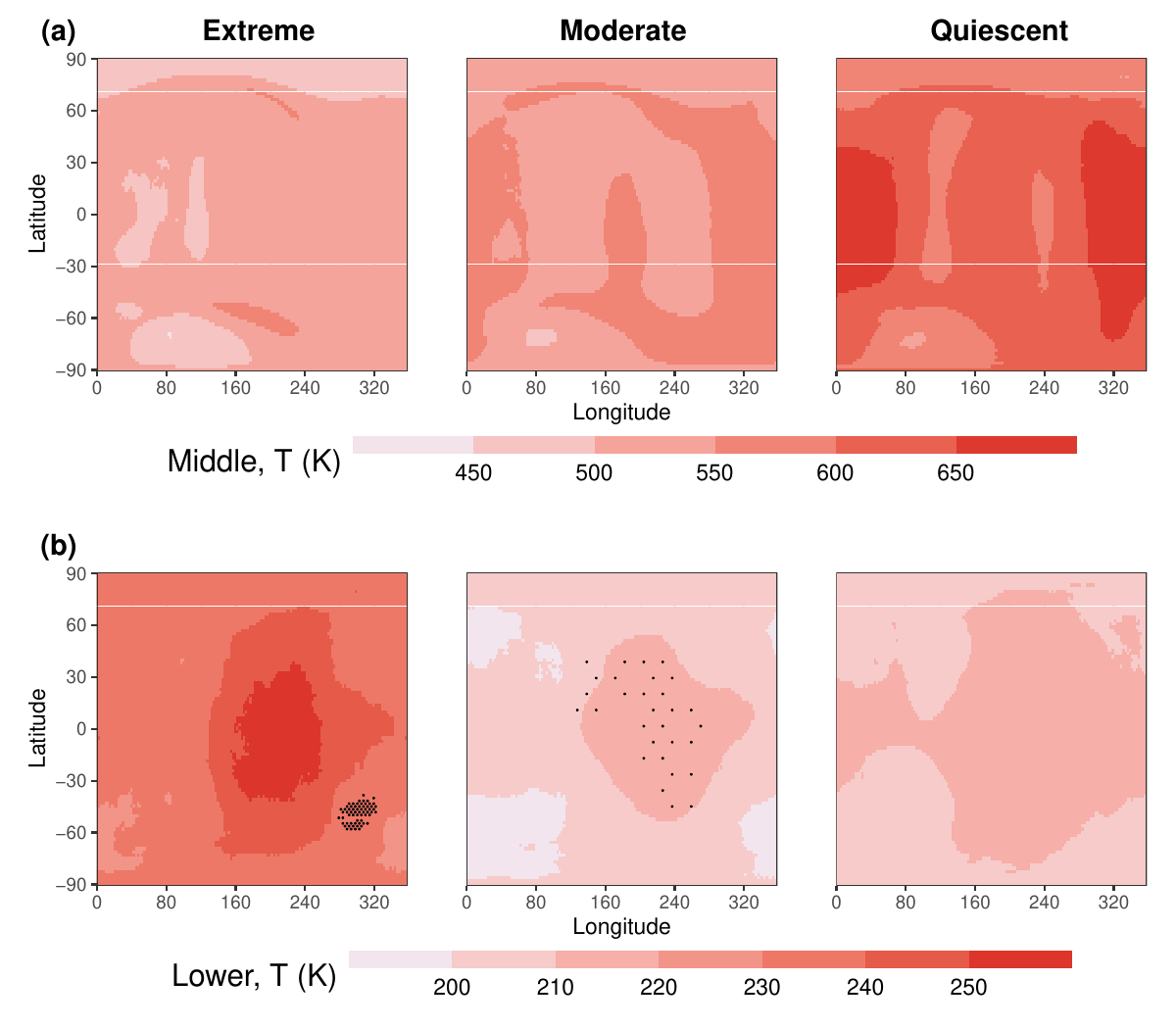}
\caption{\label{fig1} Temporal medians of 'Extreme', 'Moderate', and 'Quiescent' simulations for Temperature (T in K). Medians are taken from 8 years and for (a) $1.223\e{-5}$ mbar and (b) 0.008 mbar atmospheric levels. Extreme and Moderate medians are only taken for days when flares occur. Stippling uses a two-tailed Wilcoxon rank-sum test to represent 'Extreme' and 'Moderate' areas that are not statistically different from the 'Quiescent' one at p<0.01} 
\end{center}
\end{figure*}

\subsection{Temporal Medians}

Temporal medians of temperature for the upper and middle atmosphere in the 'Extreme', 'Active', 'Moderate', and 'Quiescent' simulations are shown in Fig.~\ref{fig1}. With the inclusion of the effects of UV flares and energetic particles, the Moderate simulation leads to 100 K of cooling in the upper atmosphere over eight years (Fig.~\ref{fig1}a). This effect is maximized with the Extreme flare cases. In the middle atmosphere regions (Fig.~\ref{fig1}b), however, the temperature response is reversed as we find warming (30-50 K) in the Extreme case. In contrast, very minimal temperature changes are found between the Quiescent and the Moderate. The peak atmospheric heating occurs at the substellar hemisphere due to the stellar UV and proton injected energies and subsequent chemical reactions that affect the distributions of nitrogen oxides and water vapor (Figs.~S\ref{figS1}-S~\ref{figS4}). We find that middle-to-lower atmospheric N$_2$O and H$_2$O, which contribute to near-surface warming, increase in the Moderate and maximize in the Extreme case. In the upper atmosphere, day-side enhancement of these species is suppressed by UV photo-dissociation of H$_2$O. Thus, an increase in the night-side is slightly more evident for H$_2$O. In the middle atmosphere, we find increased moisture across the entire planet in both Extreme and Moderate. We also see localized elevated wind speeds (Fig.~S\ref{figS2}). Nitrogen species’ mixing ratios (NO$_2$, N$_2$O) all show a steady increase as the stellar flaring activity is dialed up (Fig.~S\ref{figS3}), where O$_3$ behavior is reversed and its mixing ratio scales non-monotonically with increasing flare activity (Fig.~S\ref{figS4}). Stratospheric ozone, which influences the degree of the temperature inversion above the tropopause, reduces in its mixing ratios with increasing levels of stellar activity. These latter photochemical responses generally echo those of previous 1-D modeling results. In the middle atmosphere, mid-latitude enhancement of wind speeds and H$_2$O mixing ratios are found, with the latter particularly pronounced outside of the equatorial regions. Both types of perturbations, to regional atmospheric dynamics and water vapor abundance, may be affected by transport processes and three-dimensional atmospheric thermal structure.

\begin{figure*}[t] 
\begin{center}
\includegraphics[width=1.8\columnwidth]{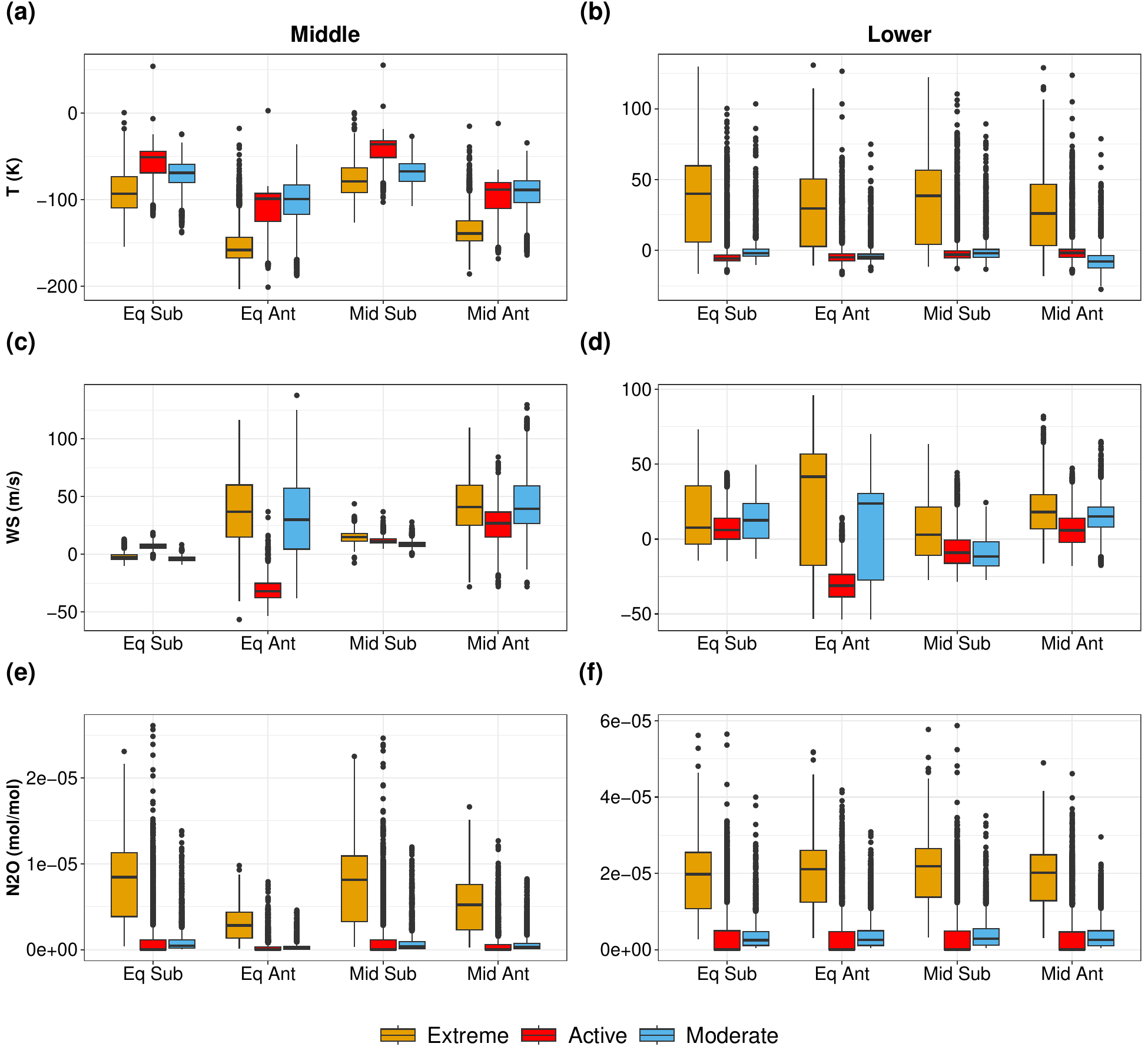}
\caption{\label{fig2} Boxplots of Extreme and Moderate anomalies for Temperature (T in K), Nitric Oxide (NO, in mol mol$^{-1}$), and Hydroxide (OH, in mol mol$^{-1}$). Anomalies are computed by subtracting the Quiescent temporal field median from the daily medians of the three sets of simulated cases: Extreme, Moderate, and Active Boxplots in (a), (c), and (e) are computed for the $1.223\e{-5}$ mbar middle atmospheric level, whereas the ones in (b), (d), and (f) represent the 0.008 mbar lower level. Each panel also represents six key regions: equatorial substellar (Eq Sub), equatorial antistellar (Eq Ant), midlatitude substellar (Mid Sub), midlatitude antistellar (Mid Ant). In all panels, at least one of the Extreme, Moderate, and Active anomaly medians significantly differ from one another at the 1\% level.} 
\end{center}
\end{figure*}

\subsection{Regional Anomalies}

Regional statistical analyses from the box and whiskers plots reveal negative temperature anomalies in the upper atmosphere (Fig. ~\ref{fig2}a). The most substantial cooling is seen in the antistellar equatorial region, where we find the greatest temperature anomalies between the Moderate and Extreme cases (Fig.~\ref{fig2}a). In the middle atmosphere, the temperature anomalies are much greater in the Extreme than in the Moderate cases (Fig.~\ref{fig2}b). All four regions of the Moderate experience minor cooling, and those of the Extreme cases experience similar magnitudes of warming. Flare-induced cooling is highest for the Extreme cases (up to 100-150 K). In the middle atmosphere (Fig. ~\ref{fig2}b), the highest (but positive) temperature anomalies are also seen in the Extreme, suggesting mesospheric heating. Simulations that experience frequent high energy flares but lowest overall number of flares (i.e., Active case) have intermediate responses in antistellar hemosphere (Fig. ~\ref{fig2}a), but the smallest thermal responses in the substellar region(s).

Chemically, except for NO mixing ratios at the middle atmosphere (Fig.~S\ref{figS5}b), NO and OH all have greater anomalies for the Extreme cases (Fig.~S\ref{figS5}a, b, e) due to enhanced day-side production rates of NO and OH driven by the proton fluxes associated with the flare events. Similar night-side NO in the Extreme and Moderate (Fig.~S\ref{figS5}a) suggests horizontal transport, whereas OH experiences little transport due to its much shorter chemical lifetime (Fig.~S\ref{figS5}e). 

The importance of flares and stellar proton events on atmospheric flow is also indicated in Fig.~\ref{fig2}c and d. The positive wind velocity anomalies across all three sets of simulations show that flare-driven dynamics, particularly in the equatorial anti-stellar and equatorial west regions for the upper and middle atmosphere, can be considerable (Fig.~\ref{fig2}c, d).

Enhancements in oxidized greenhouse gasses such as N$_2$O (Fig.~\ref{fig2}e-f) in upper and middle  atmosphere are most evident in the Extreme case. Minor effects in the other two cases. This suggests that greenhouse warming is particularly relevant for climates around the most active stars (reflected in Fig.~\ref{fig2}b). Despite having a few outliers due to sporadically large events in the Moderate and Active, on average the effects on N$_2$O are small. A similar trend is also found for effects on water vapor and hydroxyl radical (Fig.~S\ref{figS5}).


Ozone is well-studied constituent that can alter atmospheric temperature by scattering and absorbing UV radiation through the Chapman cycle. Our results show that O$_3$ abundances are sensitive to stellar flare inputs (see also \citealt{SeguraEt2010AsBio,GrenfellEt2012AsBio,scheucher2020}). The Moderate scenario shows an overall positive O$_3$ mixing ratio anomaly, while the Extreme shows a negative anomaly in the upper atmosphere (Fig.~S\ref{figS6}c). This non-linearity between input UV and the resultant ozone abundance, hinted at in the time-averaged maps in Fig.~S\ref{figS4}, is absent when examining the middle and stratospheric atmospheric abundances. For the middle atmospheric O$_3$ in particular, only negative anomalies are found in both Extreme and Moderate at the sub-stellar hemisphere, while some areas of the planet experience almost no net losses (Fig.~S\ref{figS6}d).

\begin{figure*}[t] 
\begin{center}
\includegraphics[width=1.9\columnwidth]{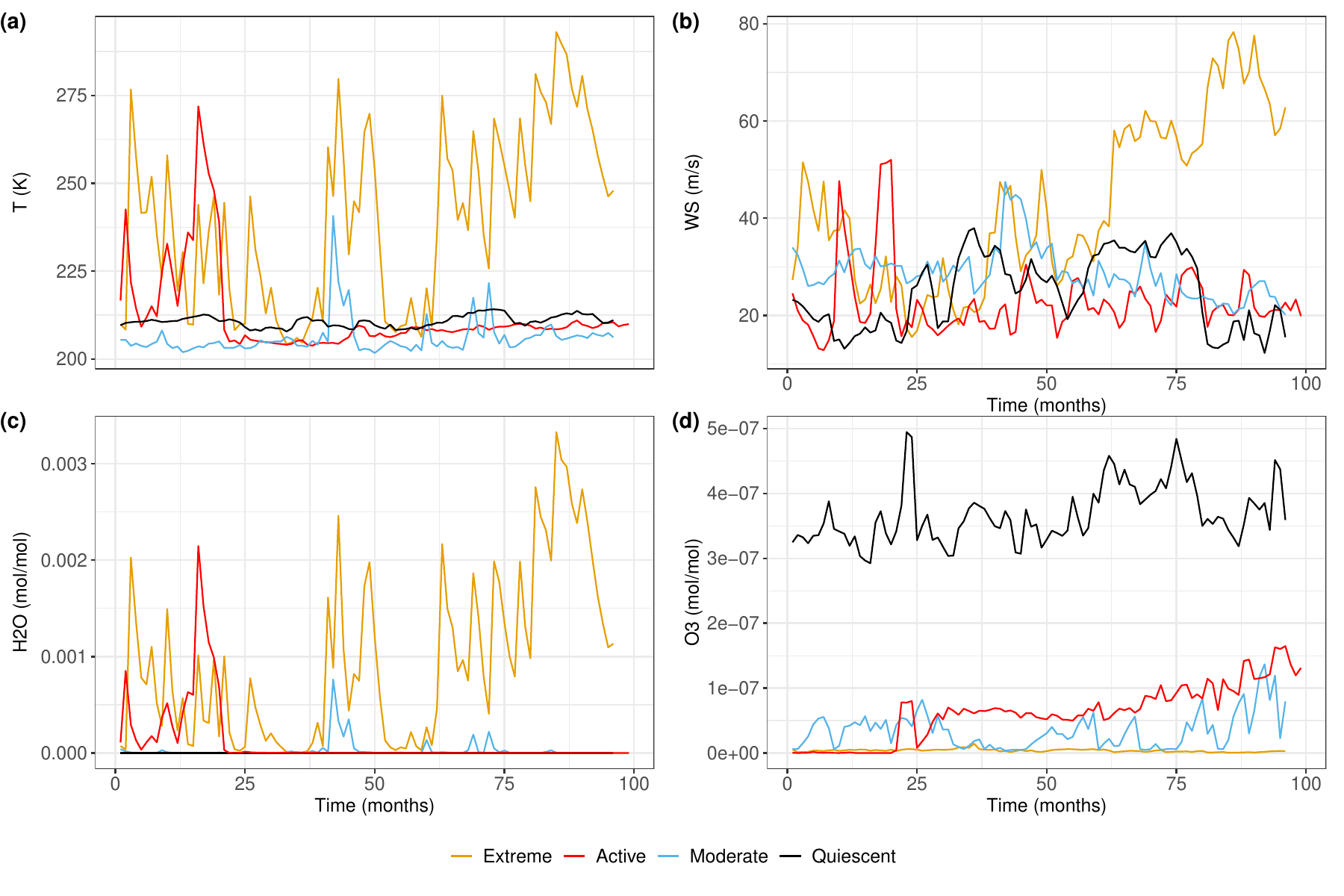}
\caption{\label{fig3} Global-median temperature (K) and horizontal wind speeds (m s$^{-1}$) at 0.008 mbar, or the middle atmosphere, showing the results for a planet around an actively flaring star (blue), around a moderately active star (red), around a quiescent/inactive star (black), and a more energetic star with a lower overall flare rate (orange). Time series are computed from the 8-year simulation.} 
\end{center}
\end{figure*}

\subsection{Time-series}

Much of the climate dynamic feedback suggests quantifying the temporal change of flare-affected atmosphereres on annual timescales may provide further insights. Fig.~\ref{fig3} shows the time series of stratospheric temperature, wind speed, water vapor mixing ratios, and ozone mixing ratios, partly driven by direct UV heating and indirectly through chemical reactions. Both NO and OH are strongly perturbed (Fig.~S\ref{figS7}), with a more spatially varied abundance seen for NO and a higher dayside concentration for OH. The chosen altitude  shows that a reversal in day–night temperature differences in the upper atmosphere can be driven by the external energy input provided by stellar events. In general, compared to the Quiescent simulations, we see that the inclusion of stellar events pushes the atmospheric states of the planet beyond that suggested by its internal climate variability alone. Peak temperatures are found to be comparable between the Extreme and the Active (Fig.~\ref{fig3}a) due to the injection of similar levels of highly energetic events. On average, however, the Extreme sees the highest temporal-mean temperature change across the eight-year period. Except for one or two temperature spikes, muted thermal oscillations are found in both the Moderate and  Active cases. 

The NO and OH abundance fluctuations can often serve as indicators of initial effect of ion pair production due to incident charged particles and UV photons accompanied by stellar events. Differences between our results and those of previous 1-D models, particularly for NO and other transport-prone constituents, are due to the 3-D nature of the CCM employed. Nitrogen oxides are first produced on the day-side hemisphere and then transported to the terminators, where night-side sustenance can be maintained. The chemical residence time of NO is around 1-2 weeks in Earth’s middle atmosphere, and based on satellite measurements, the NO decay rate towards its pre-perturbed equilibrium state is crucial to determining the atmosphere’s mean temperature change during a geomagnetic storm. Our results show that NO lifetimes are comparable to that on present Earth. Still, the relative enhancements of day versus night sides depend on the ambient condition of the atmosphere at the time of the flaring event. The cascading influence upon key molecular species, including N$_2$O, NO$_2$, H$_2$O, and O$_3$ can be seen in Fig.~\ref{fig3}c, d, and Fig.~S\ref{figS7}. 

For H$_2$O in the Moderate, only the largest events appear to disturb their mixing ratios substantially. While the effects on temperature are relatively damped for the Active case, its H$_2$O abundance change is much more considerable (Fig.~\ref{fig3}c), highlighting feedback processes even in the absence in shifts atmospheric temperature. For the Extreme case, the H$_2$O levels are at times enhanced into the moist greenhouse regime, or $>10^{-3}$ mol mol$^{-1}$ \citep{Kasting1988Icarus,Wolf+Toon2015JGR}. This enhancement is key to an exoplanet habitability and interpretation (see Sect.~\ref{sec:disc}). For the Quiescent run, the O$_3$ oscillation is tied to intrinsic variations in atmospheric temperatures and the longitudinally asymmetric stratospheric wind oscillation (LASO; \citealt{cohen2022}). In the Moderate run, such oscillatory behavior is suppressed by flares due to the mismatch of flare injection rates and the period of the LASO. In the Extreme run, traces of periodic variations in ozone abundance are largely removed due to the rapid ozone erosion through catalytic reaction cycles initiated by NO and OH.

In each of the panels for N$_2$O, NO$_2$ and H$_2$O its mixing ratio variability in the Quiescent scenario is muted or absent due to the dryness of the upper stratosphere/mesosphere and a lack of a significant N$_2$O/NO$_2$ source in the upper atmosphere of temperate planets. We also find that peak production of N$_2$O and NO$_2$ in the Extreme and Moderate simulations can be several orders of magnitude greater than planets around Quiescent stars (which lack a distinct nitrogen chemistry source in the middle atmosphere). In the Moderate run, NO$_2$ levels are not sustained by flares, while N$_2$O levels are quasi-sustained (Fig.~S\ref{figS7}a). In the Extreme run, N$_2$O levels are perturbed to a new regime where its mixing ratios fluctuate between ${\sim}2  \times 10^{-5}$ and ${\sim}3 \times 10^{-5}$ mol mol$^{-1}$, whereas NO$_2$ is rapidly returned to its pre-flare mixing ratios even after the largest of the flaring events. 

The effects of stellar flares are particularly pronounced in the middle atmosphere. At ~0.008 mbar (in the mesosphere), virtually all the observed atmospheric temperature oscillations are due to the quasi-random stellar inputs (Fig.~\ref{fig3}a). Initially, at month one but before the flaring event at month three, the atmospheric temperatures of the Extreme and Quiescent cases are similar, with the Moderate already showing some degree of cooling due to the 2-yr cutoff of our full 10-year model data (see Sect.~\ref{sec:method}). The Moderate shows much cooler temperatures throughout the period examined in this regime, with occasional temperature peaks at approximately months 40 and 73. Due to the greater assumed flare frequency and flare energies, the Extreme scenario is consistently heated by 30 to 60 K above the temperature of the Quiescent. In the final months, a complete departure from the preflare state is seen.
In contrast to the widening gap in atmospheric temperatures here, the temperature difference between the Moderate and Quiescent has been reduced throughout the simulated duration. Flaring could instigate successive cooling rather than warming episodes at other altitudes, as the earlier results indicate (see, e.g., Fig.~\ref{fig1}a). These higher levels are extremely low in their molecular densities; however, the warming effects at the stratospheric and mesospheric levels will likely have a greater impact on near-term observations than those in the upper atmosphere.

Stellar flares affect planetary wind speeds at various levels and generate interesting variability patterns not seen in climates without including time-dependent stellar inputs. In Fig.~\ref{fig3}b once again at 0.008 mbar, we find that for the 50-60 months, the flare-affected global-median wind velocities across all experiments are comparable to those generated by the natural climate variability (Quiescent curve in Fig.~\ref{fig3}b). Seen together with the temperature variations, the Extreme case, after the $\approx$63 yr mark, has constantly higher middle atmosphere temperatures and wind speeds, implying a new dynamical equilibrium is being established even when flares have turned off. On the other hand, the wind velocity variability for the Moderate is comparable with the Quiescent across the entire 8-year duration.

\begin{figure*}[t] 
\begin{center}
\includegraphics[width=1.9\columnwidth]{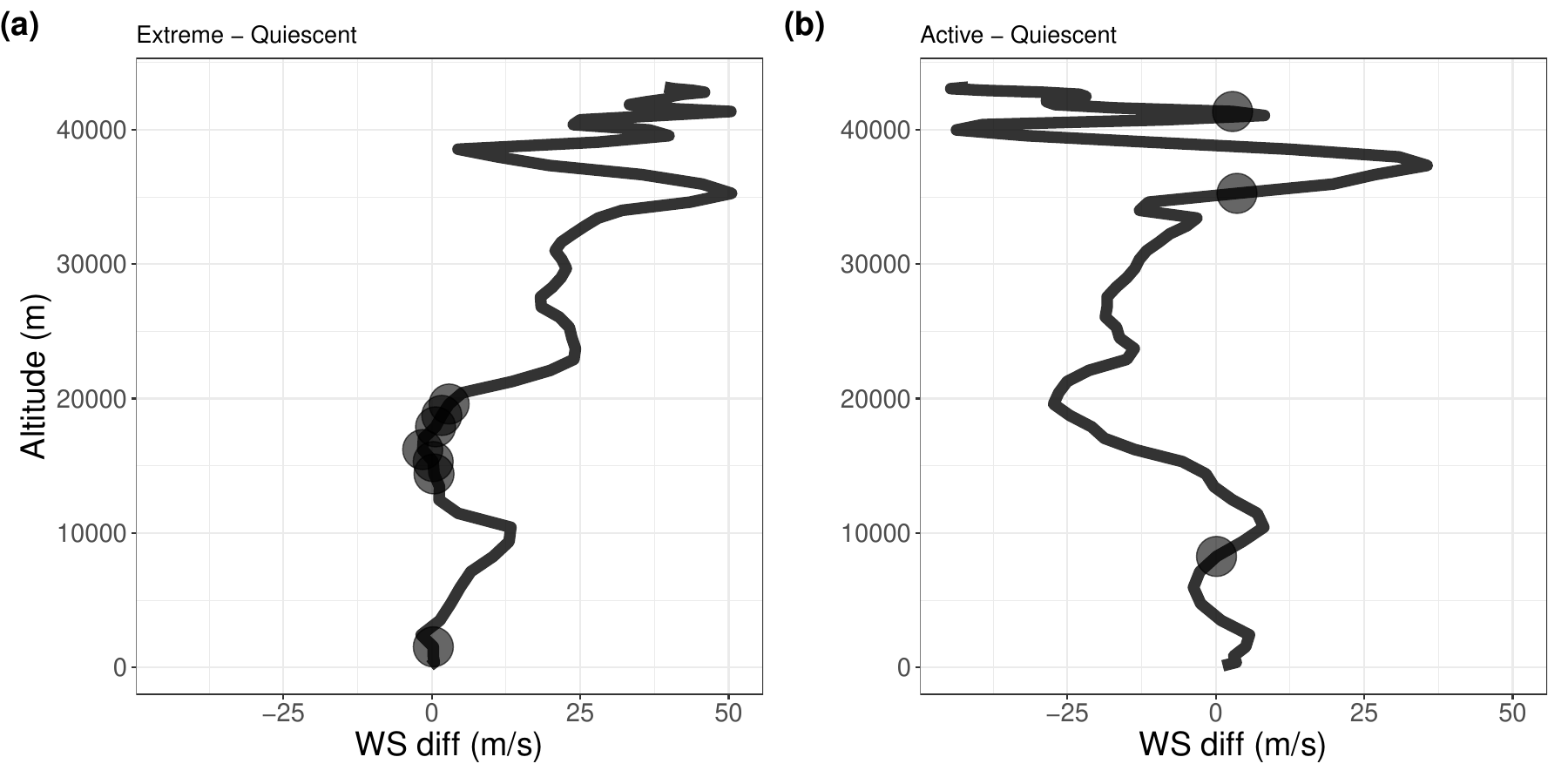}
\caption{\label{fig4} Vertical profile of wind-speed (m s$^{-1}$) temporal and field median difference between two sets of simulations over -15$^\circ$-15$^\circ$N, 0$^\circ$-50$^\circ$E. Circles represent differences that are not statistically significant at the 1\% level following a two-tailed Wilcoxon rank-sum test.} 
\end{center}
\end{figure*}

\subsection{A Glimpse into Atmospheric Dynamical Perturbations}

Cumulative and time-averaged effects are most relevant to habitability and observations. Here, we analyze the temporal-median anomalies between the Extreme and Quiescent scenarios. In the results above, simulated median wind speeds throughout the 8 years suggest that regional enhancements are pronounced (Fig.~S\ref{figS2} and Fig.~\ref{fig3}d). Wind velocity anomalies also show that flare-affected atmospheres have modulated horizontal flows that are more pronounced in specific regions of the planet. Constructing a vertical profile over longitude 0$^{\rm o}$ to 50$^{\rm o}$ and latitude -15$^{\rm o}$ to 15$^{\rm o}$, or the eastern night-side region, we find that the median horizontal WS are 25-50 m s$^{-1}$ faster between 21 and 40 km in altitude in the Extreme case (Fig.~\ref{fig4}a). Conversely, a predominant {\it decrease} in WS across its vertical profile is seen for the Active scenario, reflecting the interplay between photochemical-, and photolytic-induced temperature changes. Our results suggest that sudden increases in stellar UV and proton fluence can, at least in some regions of the planet, substantially energize the atmosphere and potentially lead to long-term asymmetries in the atmospheric dynamics on tidally-locked Earth-like exoplanets.

\vspace{4ex}

\section{Discussion \& Conclusions}
\label{sec:disc}

Observational surveys suggest that time-dependent solar and stellar activity manifested through UV flaring and Type-II bursts are common \citep{alvarado2020,veronig2021,fionnagain2022}. Previous theoretical studies investigating these effects on the temperature and dynamics of attendant atmospheres are restricted to Earth-Sun relationships. In contrast, those in the exoplanet context are mostly restricted to using one-dimensional models. Here, using a general circulation model with interactive chemistry, we study stellar flares' coupled climate and chemical effects on synchronously rotating planets with Earth-like atmospheric compositions around the host star TRAPPIST-1. In general, our photochemical results echo those of previous efforts using lower-dimensional photochemical models without including atmospheric dynamics and chemical advection. However, our 3-D modeling and analysis techniques reveal new insights on how planets around active and flaring stars could experience enhanced climate anomalies (as found by our simulated thermospheric temperature, stratospheric humidity, and wind velocity perturbations on regional and global scales). Our results suggest that, in addition to initiating key photochemical reaction pathways and inducing photochemical disequilibrium, large stellar events could affect atmospheric dynamics and even alter the planets’ circulation regime in the most extreme scenarios. For stellar environments that do not have extreme excess in X-ray or EUV irradiation, transient stellar emissions may be the dominant channel through which the dynamics of substellar atmospheres are driven. In the long term, these effects could highly alter the thermal evolution of high-mean molecular weight atmospheres, from otherwise steady climate states,  on potentially terrestrial exoplanets. The results of our study will have implications for the observations of direct-imaging missions (e.g., HWO \& LIFE)  that may be able to probe the astrophysically-influenced weather systems on habitable zone planets.

It is worth noting a few potential sources of uncertainties that may affect our results--amongst these uncertainties, two are related to the time-varying stellar inputs. The first is the potentially inaccurate UV spectra, as we replace the observed UV spectra from \citet{wilson2021} with synthetic flare spectra that are not (necessarily) representative of real TRAPPIST-1 flares or compatible with the observed optical and infrared spectra. Second, we use the same proton energy spectra for every flaring event. The 2003 Halloween geomagnetic storm is a well-documented event that may have stemmed from a strong solar flare accompanied by a coronal mass ejection. A more self-consistent approach would be using proton spectra dependent on flare characteristics (e.g., \citealt{HerbstEt2019A&Aa}). However, how the emitted proton energy of cooler, low-mass stars would scale the total flare energy for each event is still an active area of research using models of CMEs and stellar evolution \citep{hu2022,xu2024}. But that would be beyond the chief goals of this pilot study. The other two sources of uncertainty are related to the climate model itself. First, we use daily flare cadences, whereas real flares occur on timescales between minutes and hours  \citep{engebretson2008}. For larger flares, our approximation should be valid as the duration of energetic particle precipitation could persist for tens of hours to days. Conversely, higher temporal resolution would modify the effects of smaller but more frequent stellar emissions (e.g., \citealt{pettit2018effects}). Additionally, the cutoff of the atmosphere column of our climate model occurs at the lower thermosphere (or at ${\sim}10^{-6}$ mba, which is ${\sim}150$ km). Resolving the important flare-associated photo-ionization and photo-excitation processes would require at least 300-400 km model-top. Such work has already seen development for the upper atmospheres of sub-Neptune-sized planets \citep{kubyshkina2024} and for Earth-like atmospheres with 1-D simplified hydrodynamic thermosphere models \citep{tian2005}. Given the large uncertainties associated with measuring exoplanet upper atmospheres, it remains to be seen whether a full-fledged 3-D treatment is necessary for these problems (e.g., \citealt{baudino2017}).

Due to the computational expense of the 3-D model, we have limited the numerical experiments to Earth-similar N$_2$-O$_2$-dominated atmospheres, following the focus of previous works on strongly oxygenated conditions \citep{SeguraEt2010AsBio,GrenfellEt2012AsBio,TilleyEt2019AsBio}. Alternative but commonly assumed exoplanet archetypes include CO$_2$- and steam-dominated atmospheres, which will require the removal of much of the molecular oxygen that is innate to the chemical scheme (Mozart; \citealt{KinnisonEt2007JGR}) of WACCM, including the exploration of Protozoic Earth-like atmospheres (e.g., \citealt{young2024}). At the other end of the spectrum of climate archetypes are those akin to Archean Earth, or atmosphere composition in reduced redox conditions. In these conditions, one could use fully coupled climate models to understand how flaring events of the young main sequence star can influence photochemistry and aerosol production. These effects will be even more dominant due to the increased stellar activity at early times compared to older systems. At present, very few fully coupled general circulation models used to study rocky exoplanets are capable of simulating true Archean climates with the effects of photochemical haze \citep{eager2023}, while the other well-versed models for early Earth are in 1-D (see applications of such models by \citealt{ArneyEt2016AsBio,seeburger2023}).

 Flare activity can affect the surface climate evolution pathways of M-dwarf planets mainly by inducing abundance changes in key photon absorbers and scatterers. Previous work studying the early atmospheric chemistry of early Earth under a young Sun found that possible enhancement of greenhouse gases such as N$_2$O by a younger, more active Sun \citep{airapetian2016}. Our results suggest that the troposphere may experience persistent warming caused by secular flaring through a similar catalytic reaction initiated by NO production. Stellar proton events could also affect CH$_4$ abundance, another strong greenhouse gas. On the Earth, coupled climate models in conjunction with satellite observations (e.g., SABER/TIMED and MIPAS) indicate that solar storms and geomagnetic activities can drive regional changes in temperature and wind fields \citep{von2013,paivarinta2013}. In particular, \citet{becker2010} have shown that temporal heating of the middle atmosphere (90 km) can reach up to 18 K during large solar proton events, compared to our average middle atmosphere warming of 25 K (Fig.~\ref{fig3}). Another channel through which flares could affect the thermal structure is seen in the relationship between solar (geomagnetic) storms and the so-called sudden stratospheric warming (SSW). The former can generate heat ejected into the upper atmosphere, which Rossby waves could then transfer into the polar stratosphere. Other models found that the polar lower mesospheric cooling could induce an anomalous eastward flow following  the thermal wind balance. Based on these results from Earth-Sun studies, there is a need for a more detailed analysis of stellar flare-induced wave propagation with a broader range of flare behavior and flare-proton fluence relationships.

The modeled middle atmosphere effects of time-dependent stellar flaring, i.e.,modulattion of the atmospheric circulation and thermal structure over time, have implications for transmission or emission retrieval. Sufficient energetic events may even penetrate down to the lower atmosphere and affect direct imaging observations of otherwise unperturbed climates. These eruptive events may lead to vastly divergent characteristics depending on the specific nature of the event and stellar spectral type. Such suggestions have already been provided by reconstruction of the activities of the young Sun and extrapolations to other main sequence stars \citep{hu2022}. Therefore, apart from the use of a flare generator that provides light-curves and EM spectra, a better estimate of energetic particle precipitation requires improved coronal mass ejection and magnetospheric transport models \citep{fraschetti2022}. Other considerations, including flare locations and the probability for a direct ``hit", should also be included in 3D modeling techniques \citep{armitage2025}.

Other studies, using the cousin climate model WACCM6, have found the photochemical variability driven by stellar UV emissions is probably within the detection limits of future flagship direct imaging missions \citep{cooke2023a}. However, due to stellar UV uncertainties, degenerate interpretations of key spectroscopic windows would make it even more challenging for near-term missions such as the JWST \citep{cooke2023b}. Another caveat is that different sub-versions of WACCM (i.e., between versions 4-6) are expected to produce disparate results due to the different physical parameterizations used. For example, \citet{bin2018} found that distinct micro-physical and aerosol schemes within CAM4 and CAM5 can lead to different predictions in the climate states of synchronously rotating planets in the habitable zone. Given the diverse behavior of stellar emissions and the varied forms of photochemical and photolytic effects, careful model inter-comparison efforts of this sort are crucial.

The suppression, or even the complete erosion, of a substantial ozone layer is found in the Extreme simulated cases at the end of the eight-year runtime, a behavior reminiscent of earlier 1D model predictions by \citet{TilleyEt2019AsBio}. However, their study was focused on extremely long (~Gyr) timescales. As shown in our simulations, climate variability on decadal/annual timescales could be an important factor for near-term observations that would integrate over multiple (10-20) transits \citep{FauchezEt2019ApJ,lustig2019}. Our results suggest that changes in temperature and other climate factors such as wind speed are almost entirely photochemically driven with relatively little contribution from direct heating imparted by UV photons (this is in contrast to XUV-irradiated planets around young main-sequence stars, wherein the absorbed X-ray and EUV energy is transferred to thermal and mechanical energy \citep{rogers2024,johnstone2021,krissansen2024,king2024,gu2023}. Initially, climate oscillations are the most pronounced for those around active stars (e.g., Figure 3). Still, after a few years, much of their oscillatory behaviors are dampened by reduced photochemical seasonality due to the near depletion of key species such as ozone. Thus, planet atmospheres around stars with sporadic, rather than consecutive, eruptive events (which permits more time between each event for chemical species such as ozone to recover photochemically) would likely experience the greatest degrees of variability. These candidates will likely be situated around moderately-active stars. 


\section{Data Availability}
    The model data behind the figures and the scrips used to produce them can be found on\\ \href{https://zenodo.org/records/15289304}{https://zenodo.org/records/15289304}, and is also accessible via the doi link: \href{https://doi.org/10.5281/zenodo.1528930}{https://doi.org/10.5281/zenodo.15289304}.


\newpage

\bibliographystyle{apj}

\begin{thebibliography}{}
\expandafter\ifx\csname natexlab\endcsname\relax\def\natexlab#1{#1}\fi

\bibitem[{Afentakis {et~al.}(2023)Afentakis, Mullaney, Chen, Blalack, Checlair, \& Abbot}]{afentakis2023}
Afentakis, G.~P., Mullaney, K., Chen, H., {et~al.} 2023, The Astronomical Journal, 166, 117

\bibitem[{Agol {et~al.}(2021)Agol, Dorn, Grimm, Turbet, Ducrot, Delrez, Gillon, Demory, Burdanov, Barkaoui, {et~al.}}]{agol2021}
Agol, E., Dorn, C., Grimm, S.~L., {et~al.} 2021, The planetary science journal, 2, 1

\bibitem[{Airapetian {et~al.}(2016)Airapetian, Glocer, Gronoff, Hebrard, \& Danchi}]{airapetian2016}
Airapetian, V., Glocer, A., Gronoff, G., Hebrard, E., \& Danchi, W. 2016, Nature Geoscience, 9, 452

\bibitem[{Alvarado-G{\'o}mez {et~al.}(2020)Alvarado-G{\'o}mez, Drake, Fraschetti, Garraffo, Cohen, Vocks, Poppenh{\"a}ger, Moschou, Yadav, \& Manchester~IV}]{alvarado2020}
Alvarado-G{\'o}mez, J.~D., Drake, J.~J., Fraschetti, F., {et~al.} 2020, The Astrophysical Journal, 895, 47

\bibitem[{Armitage {et~al.}(2025)Armitage, Martin, \& Rodr{\'\i}guez~Mart{\'\i}nez}]{armitage2025}
Armitage, T., Martin, D.~V., \& Rodr{\'\i}guez~Mart{\'\i}nez, R. 2025, Monthly Notices of the Royal Astronomical Society, 538, 2937

\bibitem[{{Arney} {et~al.}(2016){Arney}, {Domagal-Goldman}, {Meadows}, {Wolf}, {Schwieterman}, {Charnay}, {Claire}, {H{\'e}brard}, \& {Trainer}}]{ArneyEt2016AsBio}
{Arney}, G., {Domagal-Goldman}, S.~D., {Meadows}, V.~S., {et~al.} 2016, Astrobiology, 16, 873

\bibitem[{Arsenovi{\'c} {et~al.}(2024)Arsenovi{\'c}, Rozanov, Usoskin, Turney, Sukhodolov, McCracken, Friedel, Anet, Simi{\'c}, Maliniemi, {et~al.}}]{arsenovic2024}
Arsenovi{\'c}, P., Rozanov, E., Usoskin, I., {et~al.} 2024, Proceedings of the National Academy of Sciences, 121, e2321770121

\bibitem[{Baraffe {et~al.}(2015)Baraffe, Homeier, Allard, \& Chabrier}]{baraffe2015}
Baraffe, I., Homeier, D., Allard, F., \& Chabrier, G. 2015, Astronomy \& Astrophysics, 577, A42

\bibitem[{Baudino {et~al.}(2017)Baudino, Molli{\`e}re, Venot, Tremblin, B{\'e}zard, \& Lagage}]{baudino2017}
Baudino, J.-L., Molli{\`e}re, P., Venot, O., {et~al.} 2017, The Astrophysical Journal, 850, 150

\bibitem[{Becker \& von Savigny(2010)}]{becker2010}
Becker, E., \& von Savigny, C. 2010, Journal of Geophysical Research: Atmospheres, 115

\bibitem[{Bin {et~al.}(2018)Bin, Tian, \& Liu}]{bin2018}
Bin, J., Tian, F., \& Liu, L. 2018, Earth and Planetary Science Letters, 492, 121

\bibitem[{Braam {et~al.}(2025)Braam, Palmer, Decin, Mayne, Manners, \& Rugheimer}]{braam2025}
Braam, M., Palmer, P.~I., Decin, L., {et~al.} 2025, The Planetary Science Journal, 6, 5

\bibitem[{{Carone} {et~al.}(2018){Carone}, {Keppens}, {Decin}, \& {Henning}}]{CaroneEt2018MNRAS}
{Carone}, L., {Keppens}, R., {Decin}, L., \& {Henning}, T. 2018, Monthly Notices of the Royal Astronomical Society, 473, 4672

\bibitem[{Chen {et~al.}(2023)Chen, Li, Paradise, \& Kopparapu}]{chen2023}
Chen, H., Li, G., Paradise, A., \& Kopparapu, R.~K. 2023, The Astrophysical Journal Letters, 946, L32

\bibitem[{{Chen} {et~al.}(2019){Chen}, {Wolf}, {Zhan}, \& {Horton}}]{ChenEt2019ApJ}
{Chen}, H., {Wolf}, E.~T., {Zhan}, Z., \& {Horton}, D.~E. 2019, The Astrophysical Journal, 886, 16

\bibitem[{Chen {et~al.}(2021)Chen, Zhan, Youngblood, Wolf, Feinstein, \& Horton}]{chen2021}
Chen, H., Zhan, Z., Youngblood, A., {et~al.} 2021, Nature Astronomy, 5, 298

\bibitem[{Cohen {et~al.}(2022)Cohen, Bollasina, Palmer, Sergeev, Boutle, Mayne, \& Manners}]{cohen2022}
Cohen, M., Bollasina, M.~A., Palmer, P.~I., {et~al.} 2022, The Astrophysical Journal, 930, 152

\bibitem[{Cooke {et~al.}(2023{\natexlab{a}})Cooke, Marsh, Walsh, Rugheimer, \& Villanueva}]{cooke2023a}
Cooke, G., Marsh, D., Walsh, C., Rugheimer, S., \& Villanueva, G. 2023{\natexlab{a}}, Monthly Notices of the Royal Astronomical Society, 518, 206

\bibitem[{Cooke {et~al.}(2023{\natexlab{b}})Cooke, Marsh, Walsh, \& Youngblood}]{cooke2023b}
Cooke, G., Marsh, D., Walsh, C., \& Youngblood, A. 2023{\natexlab{b}}, The Astrophysical Journal, 959, 45

\bibitem[{Davenport {et~al.}(2019)Davenport, Covey, Clarke, Boeck, Cornet, \& Hawley}]{davenport2019}
Davenport, J.~R., Covey, K.~R., Clarke, R.~W., {et~al.} 2019, The Astrophysical Journal, 871, 241

\bibitem[{{Davenport}(2016)}]{DavenportEt2016ApJ}
{Davenport}, J. R.~A. 2016, The Astrophysical Journal, 829, 23

\bibitem[{{Dong} {et~al.}(2017){Dong}, {Lingam}, {Ma}, \& {Cohen}}]{DongEt2017ApJL}
{Dong}, C., {Lingam}, M., {Ma}, Y., \& {Cohen}, O. 2017, The Astrophysical Journall, 837, L26

\bibitem[{Eager-Nash {et~al.}(2023)Eager-Nash, Mayne, Nicholson, Prins, Young, Daines, Sergeev, Lambert, Manners, Boutle, {et~al.}}]{eager2023}
Eager-Nash, J.~K., Mayne, N.~J., Nicholson, A.~E., {et~al.} 2023, Journal of Geophysical Research: Atmospheres, 128, e2022JD037544

\bibitem[{Ealy {et~al.}(2024)Ealy, Schlieder, Komacek, \& Gilbert}]{ealy2024}
Ealy, J.~N., Schlieder, J.~E., Komacek, T.~D., \& Gilbert, E.~A. 2024, The Astronomical Journal, 168, 173

\bibitem[{Engebretson {et~al.}(2008)Engebretson, Lessard, Bortnik, Green, Horne, Detrick, Weatherwax, Manninen, Petit, Posch, {et~al.}}]{engebretson2008}
Engebretson, M., Lessard, M., Bortnik, J., {et~al.} 2008, Journal of Geophysical Research: Space Physics, 113

\bibitem[{{Fauchez} {et~al.}(2019){Fauchez}, {Turbet}, {Villanueva}, {Wolf}, {Arney}, {Kopparapu}, {Lincowski}, {Mandell}, {de Wit}, {Pidhorodetska}, {Domagal-Goldman}, \& {Stevenson}}]{FauchezEt2019ApJ}
{Fauchez}, T.~J., {Turbet}, M., {Villanueva}, G.~L., {et~al.} 2019, The Astrophysical Journal, 887, 194

\bibitem[{Feinstein {et~al.}(2024)Feinstein, Seligman, France, Gagn{\'e}, \& Kowalski}]{feinstein2024}
Feinstein, A.~D., Seligman, D.~Z., France, K., Gagn{\'e}, J., \& Kowalski, A. 2024, arXiv preprint arXiv:2405.00850

\bibitem[{Fionnag{\'a}in {et~al.}(2022)Fionnag{\'a}in, Kavanagh, Vidotto, Jeffers, Petit, Marsden, Morin, Golden, Collaboration, {et~al.}}]{fionnagain2022}
Fionnag{\'a}in, D.~{\'O}., Kavanagh, R.~D., Vidotto, A.~A., {et~al.} 2022, The Astrophysical Journal, 924, 115

\bibitem[{{France} {et~al.}(2016){France}, {Loyd}, {Youngblood}, {Brown}, {Schneider}, {Hawley}, {Froning}, {Linsky}, {Roberge}, {Buccino}, {Davenport}, {Fontenla}, {Kaltenegger}, {Kowalski}, {Mauas}, {Miguel}, {Redfield}, {Rugheimer}, {Tian}, {Vieytes}, {Walkowicz}, \& {Weisenburger}}]{FranceEt2016ApJ}
{France}, K., {Loyd}, R.~O.~P., {Youngblood}, A., {et~al.} 2016, The Astrophysical Journal, 820, 89

\bibitem[{France {et~al.}(2020)France, Duvvuri, Egan, Koskinen, Wilson, Youngblood, Froning, Brown, Alvarado-G{\'o}mez, Berta-Thompson, {et~al.}}]{france2020}
France, K., Duvvuri, G., Egan, H., {et~al.} 2020, The Astronomical Journal, 160, 237

\bibitem[{Fraschetti {et~al.}(2022)Fraschetti, Alvarado-G{\'o}mez, Drake, Cohen, \& Garraffo}]{fraschetti2022}
Fraschetti, F., Alvarado-G{\'o}mez, J.~D., Drake, J.~J., Cohen, O., \& Garraffo, C. 2022, The Astrophysical Journal, 937, 126

\bibitem[{Fromont {et~al.}(2024)Fromont, Ahlers, do~Amaral, Barnes, Gilbert, Quintana, Peacock, Barclay, \& Youngblood}]{fromont2024}
Fromont, E.~F., Ahlers, J.~P., do~Amaral, L.~N., {et~al.} 2024, The Astrophysical Journal, 961, 115

\bibitem[{{Funke} {et~al.}(2011){Funke}, {Baumgaertner}, {Calisto}, {Egorova}, {Jackman}, {Kieser}, {Krivolutsky}, {L{\'o}pez-Puertas}, {Marsh}, {Reddmann}, {Rozanov}, {Salmi}, {Sinnhuber}, {Stiller}, {Verronen}, {Versick}, {von Clarmann}, {Vyushkova}, {Wieters}, \& {Wissing}}]{Funke:2011}
{Funke}, B., {Baumgaertner}, A., {Calisto}, M., {et~al.} 2011, Atmospheric Chemistry \& Physics, 11, 9089

\bibitem[{{Grenfell} {et~al.}(2012){Grenfell}, {Grie{\ss}meier}, {von Paris}, {Patzer}, {Lammer}, {Stracke}, {Gebauer}, {Schreier}, \& {Rauer}}]{GrenfellEt2012AsBio}
{Grenfell}, J.~L., {Grie{\ss}meier}, J.-M., {von Paris}, P., {et~al.} 2012, Astrobiology, 12, 1109

\bibitem[{Gu \& Chen(2023)}]{gu2023}
Gu, P.-G., \& Chen, H. 2023, The Astrophysical Journal Letters, 953, L27

\bibitem[{{Hack}(1994)}]{Hack1994JGR}
{Hack}, J.~J. 1994, Journal of Geophysical Research, 99, 5551

\bibitem[{He {et~al.}(2022)He, Merrelli, L’Ecuyer, \& Turnbull}]{he2022}
He, F., Merrelli, A., L’Ecuyer, T.~S., \& Turnbull, M.~C. 2022, The Astrophysical Journal, 933, 62

\bibitem[{{Herbst} {et~al.}(2019){Herbst}, {Papaioannou}, {Banjac}, \& {Heber}}]{HerbstEt2019A&Aa}
{Herbst}, K., {Papaioannou}, A., {Banjac}, S., \& {Heber}, B. 2019, Astronomy \& Astrophysics, 621, A67

\bibitem[{Howard \& MacGregor(2022)}]{howard2022}
Howard, W.~S., \& MacGregor, M.~A. 2022, The Astrophysical Journal, 926, 204

\bibitem[{Howard {et~al.}(2020)Howard, Corbett, Law, Ratzloff, Galliher, Glazier, Gonzalez, Soto, Fors, Del~Ser, {et~al.}}]{howard2020}
Howard, W.~S., Corbett, H., Law, N.~M., {et~al.} 2020, The Astrophysical Journal, 902, 115

\bibitem[{Howard {et~al.}(2024)Howard, MacGregor, Feinstein, Vega, Cody, Turner, Scott, Burt, \& Venuti}]{howard2024}
Howard, W.~S., MacGregor, M.~A., Feinstein, A.~D., {et~al.} 2024, The Astronomical Journal, 169, 27

\bibitem[{Hu {et~al.}(2022)Hu, Airapetian, Li, Zank, \& Jin}]{hu2022}
Hu, J., Airapetian, V.~S., Li, G., Zank, G., \& Jin, M. 2022, Science Advances, 8, eabi9743

\bibitem[{Ilin {et~al.}(2021)Ilin, Schmidt, Poppenh{\"a}ger, Davenport, Kristiansen, \& Omohundro}]{ilin2021}
Ilin, E., Schmidt, S.~J., Poppenh{\"a}ger, K., {et~al.} 2021, Astronomy \& Astrophysics, 645, A42

\bibitem[{Jackman {et~al.}(2008)Jackman, Marsh, Vitt, Garcia, Fleming, Labow, Randall, L{\'o}pez-Puertas, Funke, Clarmann, {et~al.}}]{jackman2008short}
Jackman, C., Marsh, D., Vitt, F., {et~al.} 2008, Atmospheric Chemistry and Physics, 8, 765

\bibitem[{Jackman {et~al.}(2023)Jackman, Shkolnik, Million, Fleming, Richey-Yowell, \& Loyd}]{jackman2023}
Jackman, J.~A., Shkolnik, E.~L., Million, C., {et~al.} 2023, Monthly Notices of the Royal Astronomical Society, 519, 3564

\bibitem[{Johnstone {et~al.}(2021)Johnstone, Lammer, Kislyakova, Scherf, \& G{\"u}del}]{johnstone2021}
Johnstone, C.~P., Lammer, H., Kislyakova, K.~G., Scherf, M., \& G{\"u}del, M. 2021, Earth and Planetary Science Letters, 576, 117197

\bibitem[{{Kasting}(1988)}]{Kasting1988Icarus}
{Kasting}, J.~F. 1988, Icarusus, 74, 472

\bibitem[{Kiehl \& Ramanathan(1983)}]{kiehl1983co2}
Kiehl, J., \& Ramanathan, V. 1983, Journal of Geophysical Research: Oceans, 88, 5191

\bibitem[{King {et~al.}(2024)King, Corrales, Fern{\'a}ndez~Fern{\'a}ndez, Wheatley, Malsky, Osborn, \& Armstrong}]{king2024}
King, G.~W., Corrales, L.~R., Fern{\'a}ndez~Fern{\'a}ndez, J., {et~al.} 2024, Monthly Notices of the Royal Astronomical Society, 530, 3500

\bibitem[{{Kinnison} {et~al.}(2007){Kinnison}, {Brasseur}, {Walters}, {Garcia}, {Marsh}, {Sassi}, {Harvey}, {Randall}, {Emmons}, {Lamarque}, {Hess}, {Orlando}, {Tie}, {Randel}, {Pan}, {Gettelman}, {Granier}, {Diehl}, {Niemeier}, \& {Simmons}}]{KinnisonEt2007JGR}
{Kinnison}, D.~E., {Brasseur}, G.~P., {Walters}, S., {et~al.} 2007, Journal of Geophysical Research (Atmospheres), 112, D20302

\bibitem[{Komacek \& Abbot(2019)}]{komacek2019}
Komacek, T.~D., \& Abbot, D.~S. 2019, The Astrophysical Journal, 871, 245

\bibitem[{Kopparapu {et~al.}(2017)Kopparapu, Wolf, Arney, Batalha, Haqq-Misra, Grimm, \& Heng}]{kumar2017}
Kopparapu, R., Wolf, E.~T., Arney, G., {et~al.} 2017, The Astrophysical Journal, 845, 5

\bibitem[{Kopparapu {et~al.}(2016)Kopparapu, Wolf, Haqq-Misra, Yang, Kasting, Meadows, Terrien, \& Mahadevan}]{kumar2016inner}
Kopparapu, R., Wolf, E.~T., Haqq-Misra, J., {et~al.} 2016, The Astrophysical Journal, 819, 84

\bibitem[{Kowalski {et~al.}(2024{\natexlab{a}})Kowalski, Allred, \& Carlsson}]{kowalski2024time}
Kowalski, A.~F., Allred, J.~C., \& Carlsson, M. 2024{\natexlab{a}}, arXiv preprint arXiv:2404.13214

\bibitem[{{Kowalski} {et~al.}(2013){Kowalski}, {Hawley}, {Wisniewski}, {Osten}, {Hilton}, {Holtzman}, {Schmidt}, \& {Davenport}}]{KowalskiEt2013ApJS}
{Kowalski}, A.~F., {Hawley}, S.~L., {Wisniewski}, J.~P., {et~al.} 2013, The Astrophysical Journal Supplement Series, 207, 15

\bibitem[{Kowalski {et~al.}(2024{\natexlab{b}})Kowalski, Osten, Notsu, Tristan, Segura, Maehara, Namekata, \& Inoue}]{kowalski2024}
Kowalski, A.~F., Osten, R.~A., Notsu, Y., {et~al.} 2024{\natexlab{b}}, The Astrophysical Journal, 978, 81

\bibitem[{Krissansen-Totton {et~al.}(2024)Krissansen-Totton, Wogan, Thompson, \& Fortney}]{krissansen2024}
Krissansen-Totton, J., Wogan, N., Thompson, M., \& Fortney, J.~J. 2024, Nature communications, 15, 8374

\bibitem[{Kruskal \& Wallis(1952)}]{kruskal1952}
Kruskal, W.~H., \& Wallis, W.~A. 1952, Journal of the American statistical Association, 47, 583

\bibitem[{Kubyshkina {et~al.}(2024)Kubyshkina, Fossati, \& Erkaev}]{kubyshkina2024}
Kubyshkina, D., Fossati, L., \& Erkaev, N.~V. 2024, Astronomy \& Astrophysics, 684, A26

\bibitem[{{L{\'o}pez-Puertas} {et~al.}(2005){L{\'o}pez-Puertas}, {Funke}, {Gil-L{\'o}pez}, {von Clarmann}, {Stiller}, {H{\"o}Pfner}, {Kellmann}, {Fischer}, \& {Jackman}}]{LopezEt2005JGR}
{L{\'o}pez-Puertas}, M., {Funke}, B., {Gil-L{\'o}pez}, S., {et~al.} 2005, Journal of Geophysical Research (Space Physics), 110, A09S43

\bibitem[{Louca {et~al.}(2023)Louca, Miguel, Tsai, Froning, Loyd, \& France}]{louca2023}
Louca, A.~J., Miguel, Y., Tsai, S.-M., {et~al.} 2023, Monthly Notices of the Royal Astronomical Society, 521, 3333

\bibitem[{{Loyd} {et~al.}(2018){Loyd}, {France}, {Youngblood}, {Schneider}, {Brown}, {Hu}, {Segura}, {Linsky}, {Redfield}, {Tian}, {Rugheimer}, {Miguel}, \& {Froning}}]{LoydEt2018ApJa}
{Loyd}, R.~O.~P., {France}, K., {Youngblood}, A., {et~al.} 2018, The Astrophysical Journal, 867, 71

\bibitem[{Loyd {et~al.}(2018)Loyd, Shkolnik, Schneider, Barman, Meadows, Pagano, \& Peacock}]{loyd2018}
Loyd, R.~P., Shkolnik, E.~L., Schneider, A.~C., {et~al.} 2018, The Astrophysical Journal, 867, 70

\bibitem[{Luo {et~al.}(2023)Luo, Hu, Yang, Zhang, \& Yung}]{luo2023}
Luo, Y., Hu, Y., Yang, J., Zhang, M., \& Yung, Y.~L. 2023, Proceedings of the National Academy of Sciences, 120, e2309312120

\bibitem[{Lustig-Yaeger {et~al.}(2019)Lustig-Yaeger, Meadows, \& Lincowski}]{lustig2019}
Lustig-Yaeger, J., Meadows, V.~S., \& Lincowski, A.~P. 2019, The Astronomical Journal, 158, 27

\bibitem[{Mann \& Whitney(1947)}]{mann1947}
Mann, H.~B., \& Whitney, D.~R. 1947, The annals of mathematical statistics, 50

\bibitem[{{Marsh} {et~al.}(2013){Marsh}, {Mills}, {Kinnison}, {Lamarque}, {Calvo}, \& {Polvani}}]{MarshEt2013JGR}
{Marsh}, D.~R., {Mills}, M.~J., {Kinnison}, D.~E., {et~al.} 2013, Journal of Climate, 26, 7372

\bibitem[{Merlis \& Schneider(2010)}]{merlis2010}
Merlis, T.~M., \& Schneider, T. 2010, Journal of Advances in Modeling Earth Systems, 2

\bibitem[{Neale {et~al.}(2010)Neale, Chen, Gettelman, Lauritzen, Park, Williamson, Conley, Garcia, Kinnison, Lamarque, {et~al.}}]{neale2010description}
Neale, R.~B., Chen, C.-C., Gettelman, A., {et~al.} 2010, NCAR Tech. Note NCAR/TN-486+ STR

\bibitem[{Osten {et~al.}(2016)Osten, Kowalski, Drake, Krimm, Page, Gazeas, Kennea, Oates, Page, De~Miguel, {et~al.}}]{osten2016}
Osten, R.~A., Kowalski, A., Drake, S.~A., {et~al.} 2016, The Astrophysical Journal, 832, 174

\bibitem[{P{\"a}iv{\"a}rinta {et~al.}(2013)P{\"a}iv{\"a}rinta, Sepp{\"a}l{\"a}, Andersson, Verronen, Th{\"o}lix, \& Kyr{\"o}l{\"a}}]{paivarinta2013}
P{\"a}iv{\"a}rinta, S.-M., Sepp{\"a}l{\"a}, A., Andersson, M., {et~al.} 2013, Journal of Geophysical Research: Atmospheres, 118, 6837

\bibitem[{Paudel {et~al.}(2024)Paudel, Barclay, Youngblood, Quintana, Schlieder, Vega, Gilbert, Osten, Peacock, Tristan, {et~al.}}]{paudel2024}
Paudel, R.~R., Barclay, T., Youngblood, A., {et~al.} 2024, The Astrophysical Journal, 971, 24

\bibitem[{Pettit {et~al.}(2018)Pettit, Randall, Marsh, Bardeen, Qian, Jackman, Woods, Coster, \& Harvey}]{pettit2018effects}
Pettit, J., Randall, C., Marsh, D., {et~al.} 2018, Journal of Geophysical Research: Space Physics

\bibitem[{Ridgway {et~al.}(2023)Ridgway, Zamyatina, Mayne, Manners, Lambert, Braam, Drummond, H{\'e}brard, Palmer, \& Kohary}]{ridgway2023}
Ridgway, R.~J., Zamyatina, M., Mayne, N.~J., {et~al.} 2023, Monthly Notices of the Royal Astronomical Society, 518, 2472

\bibitem[{Rogers {et~al.}(2024)Rogers, Owen, \& Schlichting}]{rogers2024}
Rogers, J.~G., Owen, J.~E., \& Schlichting, H.~E. 2024, Monthly Notices of the Royal Astronomical Society, 529, 2716

\bibitem[{Rozanov {et~al.}(2005)Rozanov, Callis, Schlesinger, Yang, Andronova, \& Zubov}]{rozanov2005}
Rozanov, E., Callis, L., Schlesinger, M., {et~al.} 2005, Geophysical Research Letters, 32

\bibitem[{Rustamkulov {et~al.}(2023)Rustamkulov, Sing, Mukherjee, May, Kirk, Schlawin, Line, Piaulet, Carter, Batalha, {et~al.}}]{rustamkulov2023}
Rustamkulov, Z., Sing, D., Mukherjee, S., {et~al.} 2023, Nature, 614, 659

\bibitem[{Sagan \& Mullen(1972)}]{sagan1972}
Sagan, C., \& Mullen, G. 1972, Science, 177, 52

\bibitem[{Scheucher {et~al.}(2020)Scheucher, Herbst, Schmidt, Grenfell, Schreier, Banjac, Heber, Rauer, \& Sinnhuber}]{scheucher2020}
Scheucher, M., Herbst, K., Schmidt, V., {et~al.} 2020, The Astrophysical Journal, 893, 12

\bibitem[{Seeburger {et~al.}(2023)Seeburger, Higgins, Whiteford, \& Cockell}]{seeburger2023}
Seeburger, R., Higgins, P.~M., Whiteford, N.~P., \& Cockell, C.~S. 2023, Astrobiology, 23, 415

\bibitem[{{Segura} {et~al.}(2010){Segura}, {Walkowicz}, {Meadows}, {Kasting}, \& {Hawley}}]{SeguraEt2010AsBio}
{Segura}, A., {Walkowicz}, L.~M., {Meadows}, V., {Kasting}, J., \& {Hawley}, S. 2010, Astrobiology, 10, 751

\bibitem[{Sergeev {et~al.}(2020)Sergeev, Lambert, Mayne, Boutle, Manners, \& Kohary}]{sergeev2020}
Sergeev, D.~E., Lambert, F.~H., Mayne, N.~J., {et~al.} 2020, The Astrophysical Journal, 894, 84

\bibitem[{Stelzer {et~al.}(2022)Stelzer, Caramazza, Raetz, Argiroffi, \& Coffaro}]{stelzer2022}
Stelzer, B., Caramazza, M., Raetz, S., Argiroffi, C., \& Coffaro, M. 2022, Astronomy \& Astrophysics, 667, L9

\bibitem[{Sukhodolov {et~al.}(2017)Sukhodolov, Usoskin, Rozanov, Asvestari, Ball, Curran, Fischer, Kovaltsov, Miyake, Peter, {et~al.}}]{sukhodolov2017}
Sukhodolov, T., Usoskin, I., Rozanov, E., {et~al.} 2017, Scientific Reports, 7, 45257

\bibitem[{Tian {et~al.}(2005)Tian, Toon, Pavlov, \& De~Sterck}]{tian2005}
Tian, F., Toon, O.~B., Pavlov, A.~A., \& De~Sterck, H. 2005, The Astrophysical Journal, 621, 1049

\bibitem[{{Tilley} {et~al.}(2019){Tilley}, {Segura}, {Meadows}, {Hawley}, \& {Davenport}}]{TilleyEt2019AsBio}
{Tilley}, M.~A., {Segura}, A., {Meadows}, V., {Hawley}, S., \& {Davenport}, J. 2019, Astrobiology, 19, 64

\bibitem[{Tsai {et~al.}(2023)Tsai, Lee, Powell, Gao, Zhang, Moses, H{\'e}brard, Venot, Parmentier, Jordan, {et~al.}}]{tsai2023}
Tsai, S.-M., Lee, E.~K., Powell, D., {et~al.} 2023, Nature, 617, 483

\bibitem[{Turbet {et~al.}(2016)Turbet, Leconte, Selsis, Bolmont, Forget, Ribas, Raymond, \& Anglada-Escud{\'e}}]{turbet2016}
Turbet, M., Leconte, J., Selsis, F., {et~al.} 2016, Astronomy \& Astrophysics, 596, A112

\bibitem[{Van~Doorsselaere {et~al.}(2017)Van~Doorsselaere, Shariati, \& Debosscher}]{van2017}
Van~Doorsselaere, T., Shariati, H., \& Debosscher, J. 2017, The Astrophysical Journal Supplement Series, 232, 26

\bibitem[{Veronig {et~al.}(2021)Veronig, Odert, Leitzinger, Dissauer, Fleck, \& Hudson}]{veronig2021}
Veronig, A.~M., Odert, P., Leitzinger, M., {et~al.} 2021, Nature Astronomy, 5, 697

\bibitem[{Vida {et~al.}(2021)Vida, B{\'o}di, Szklen{\'a}r, \& Seli}]{vida2021}
Vida, K., B{\'o}di, A., Szklen{\'a}r, T., \& Seli, B. 2021, Astronomy \& Astrophysics, 652, A107

\bibitem[{Von~Clarmann {et~al.}(2013)Von~Clarmann, Funke, L{\'o}pez-Puertas, Kellmann, Linden, Stiller, Jackman, \& Harvey}]{von2013}
Von~Clarmann, T., Funke, B., L{\'o}pez-Puertas, M., {et~al.} 2013, Geophysical Research Letters, 40, 2339

\bibitem[{Way \& Del~Genio(2020)}]{way2020}
Way, M.~J., \& Del~Genio, A.~D. 2020, Journal of Geophysical Research: Planets, 125, e2019JE006276

\bibitem[{{Way} {et~al.}(2016){Way}, {Del Genio}, {Kiang}, {Sohl}, {Grinspoon}, {Aleinov}, {Kelley}, \& {Clune}}]{WayEt2016GRL}
{Way}, M.~J., {Del Genio}, A.~D., {Kiang}, N.~Y., {et~al.} 2016, Geophysical Research Letters, 43, 8376

\bibitem[{Wilson {et~al.}(2021)Wilson, Froning, Duvvuri, France, Youngblood, Schneider, Berta-Thompson, Brown, Buccino, Hawley, {et~al.}}]{wilson2021}
Wilson, D.~J., Froning, C.~S., Duvvuri, G.~M., {et~al.} 2021, The Astrophysical Journal, 911, 18

\bibitem[{{Wolf} \& {Toon}(2015)}]{Wolf+Toon2015JGR}
{Wolf}, E.~T., \& {Toon}, O.~B. 2015, Journal of Geophysical Research (Atmospheres), 120, 5775

\bibitem[{Wordsworth \& Pierrehumbert(2014)}]{WordsworthEt2014ApJL}
Wordsworth, R., \& Pierrehumbert, R. 2014, The Astrophysical Journal Letters, 785, L20

\bibitem[{Xu {et~al.}(2024)Xu, Alvarado-G{\'o}mez, Tian, Poppenh{\"a}ger, Guerrero, \& Liu}]{xu2024}
Xu, Y., Alvarado-G{\'o}mez, J.~D., Tian, H., {et~al.} 2024, The Astrophysical Journal, 971, 153

\bibitem[{Yang {et~al.}(2013)Yang, Cowan, \& Abbot}]{yang2013stabilizing}
Yang, J., Cowan, N.~B., \& Abbot, D.~S. 2013, The Astrophysical Journal Letters, 771, L45

\bibitem[{Young {et~al.}(2024)Young, Robinson, Krissansen-Totton, Schwieterman, Wogan, Way, Sohl, Arney, Reinhard, Line, {et~al.}}]{young2024}
Young, A.~V., Robinson, T.~D., Krissansen-Totton, J., {et~al.} 2024, Nature Astronomy, 8, 101

\bibitem[{{Youngblood} {et~al.}(2017){Youngblood}, {France}, {Loyd}, {Brown}, {Mason}, {Schneider}, {Tilley}, {Berta-Thompson}, {Buccino}, {Froning}, {Hawley}, {Linsky}, {Mauas}, {Redfield}, {Kowalski}, {Miguel}, {Newton}, {Rugheimer}, {Segura}, {Roberge}, \& {Vieytes}}]{YoungbloodEt2017ApJ}
{Youngblood}, A., {France}, K., {Loyd}, R.~O.~P., {et~al.} 2017, The Astrophysical Journal, 843, 31

\bibitem[{Zhang \& McFarlane(1995)}]{zhang1995sensitivity}
Zhang, G.~J., \& McFarlane, N.~A. 1995, Atmosphere-ocean, 33, 407

\bibitem[{Zhang {et~al.}(2003)Zhang, Lin, Bretherton, Hack, \& Rasch}]{zhang2003modified}
Zhang, M., Lin, W., Bretherton, C.~S., Hack, J.~J., \& Rasch, P.~J. 2003, Journal of Geophysical Research: Atmospheres, 108, ACL

\end{thebibliography}

\appendix

Figures in appendix 1-7 contain results that support our findings in the main text.
\renewcommand{\figurename}{Appendix Figure}
\setcounter{figure}{0}
\begin{figure*}[h] 
\begin{center}
\includegraphics[width=.8\columnwidth]{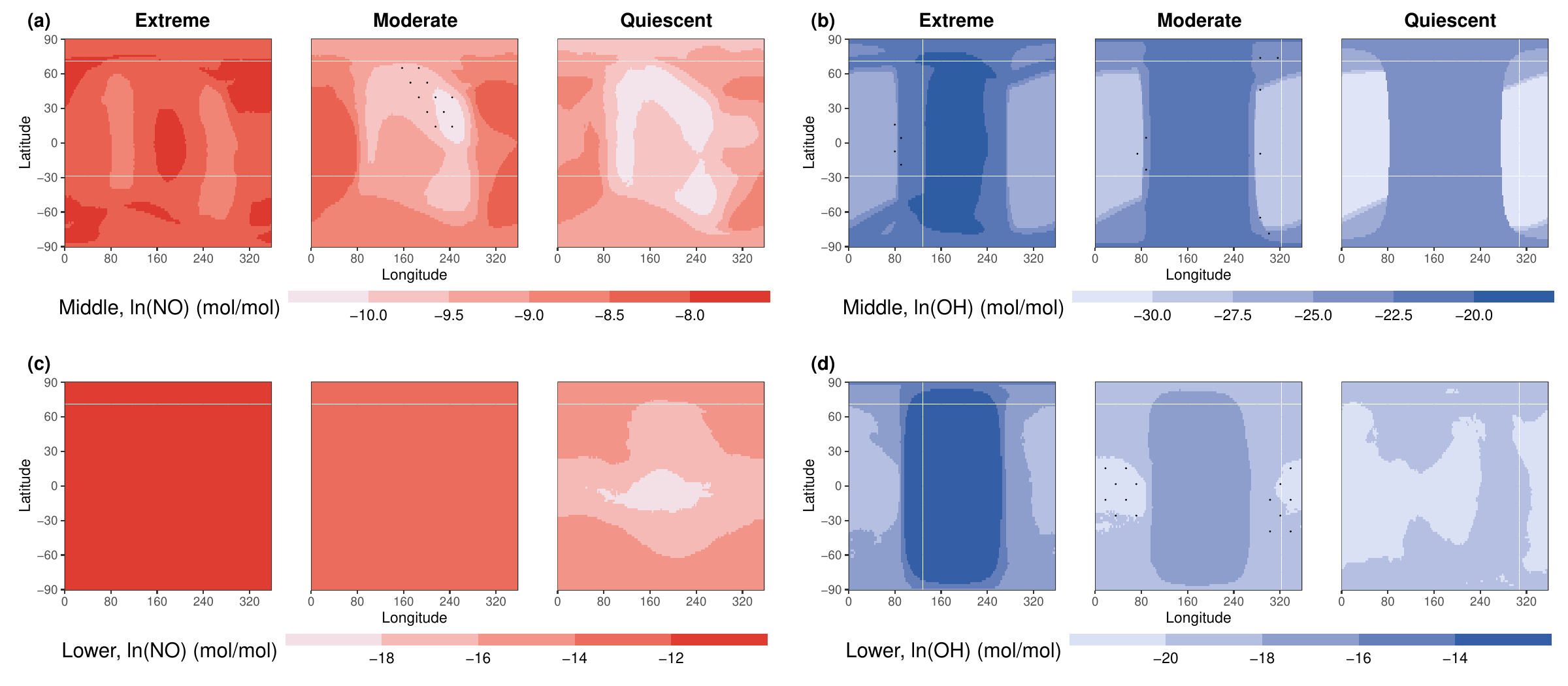}
\caption{\label{figS1} Same as Figure 1, but for Nitrogen Oxide (NO in mol mol$^{-1}$) and Hydroxide (OH in mol mol$^{-1}$). Given the large spread between the simulations, values are shown as natural logarithms. } 
\end{center}
\end{figure*}  

\begin{figure*}[h] 
\begin{center}
\includegraphics[width=.8\columnwidth]{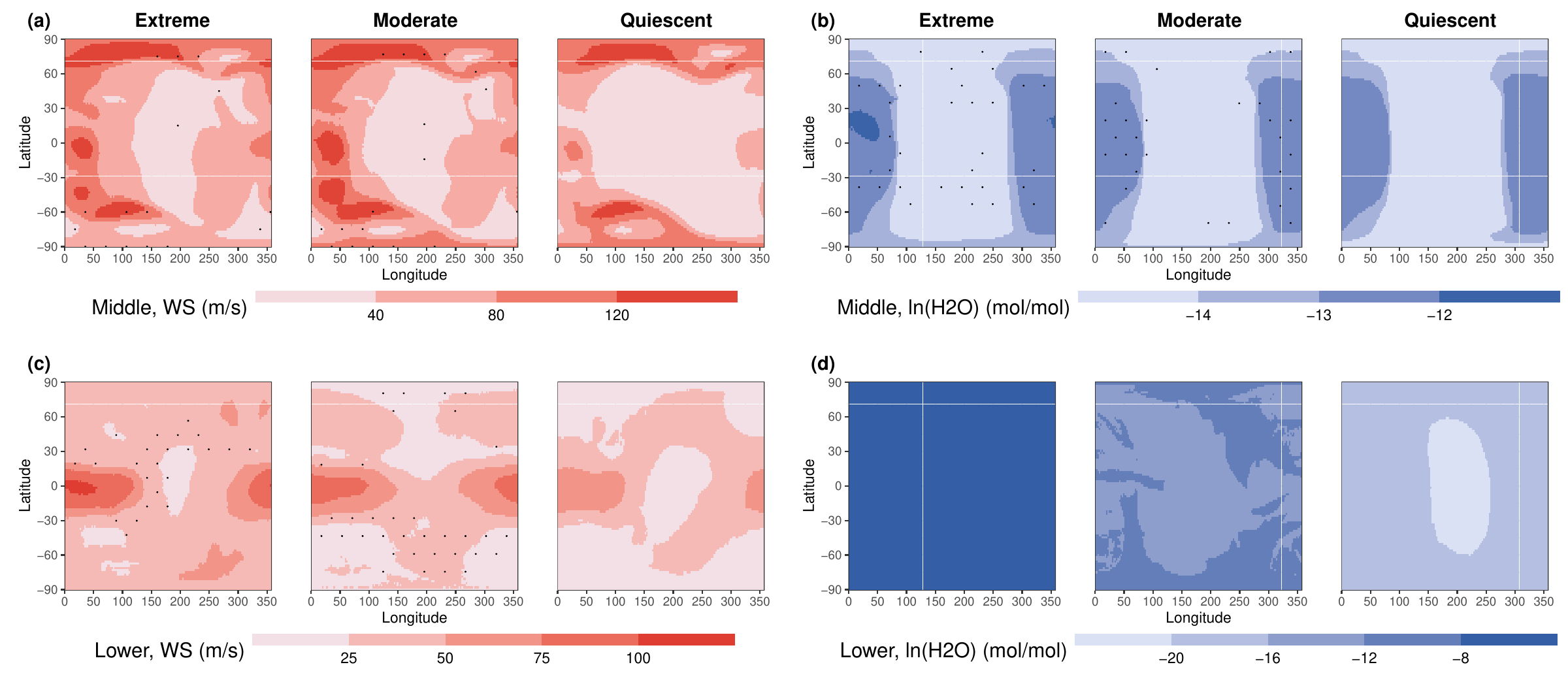}
\caption{\label{figS2} Same as Figure 1, but for Wind Speed (WS, m s$^{-1}$) and Water Vapour (H$_2$O, mol mol$^{-1}$) as natural logarithms.} 
\end{center}
\end{figure*}  

\begin{figure*}[h] 
\begin{center}
\includegraphics[width=.8\columnwidth]{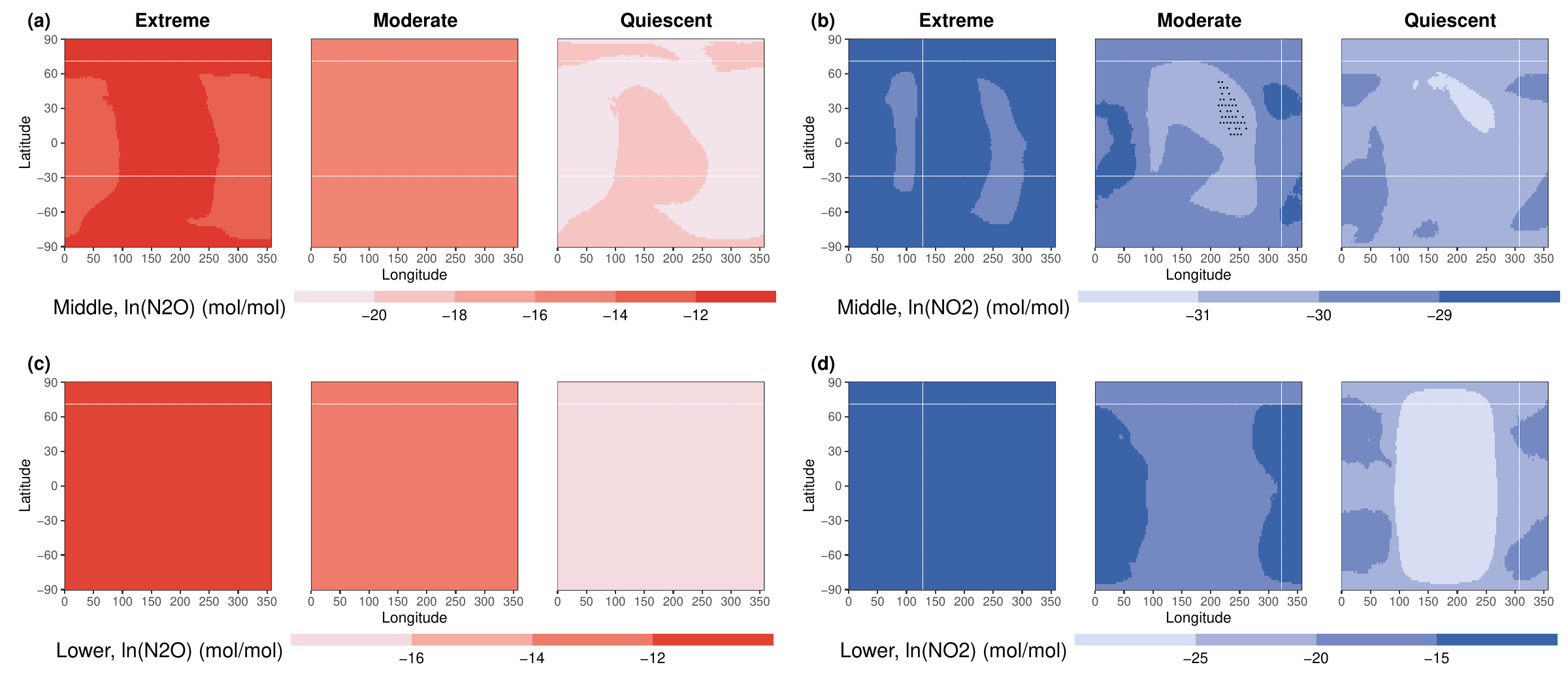}
\caption{\label{figS3} Same as Figure 2 but for Nitrous Oxide (N$_2$O) and Nitrogen Dioxide (NO$_2$, mol mol$^{-1}$) as natural logarithms.} 
\end{center}
\end{figure*}  

\begin{figure*}[h] 
\begin{center}
\includegraphics[width=.8\columnwidth]{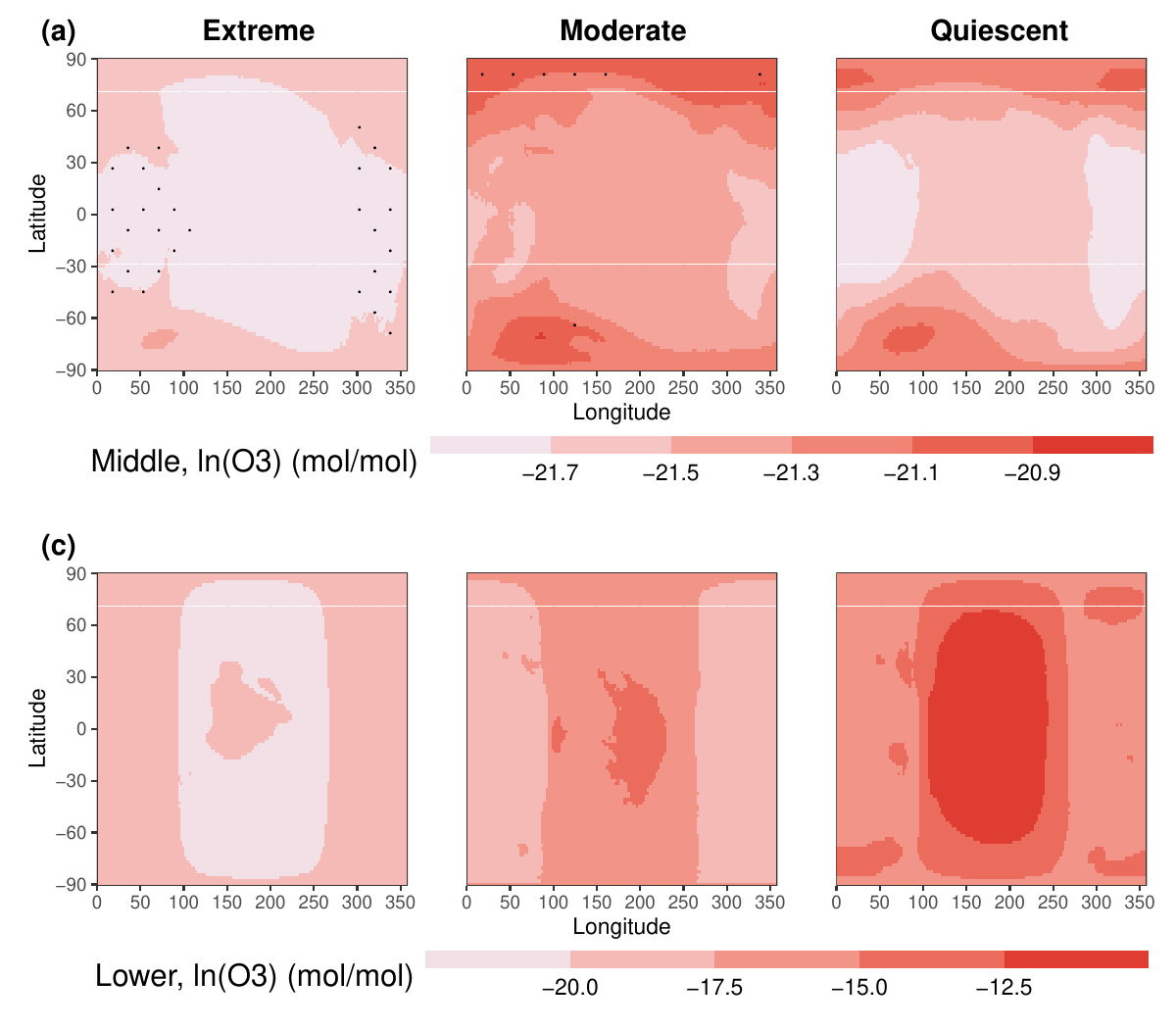}
\caption{\label{figS4} Same as Figure 1, but for Ozone (O$_3$, mol mol$^{-1}$) in natural logarithm.} 
\end{center}

\end{figure*}  
\begin{figure*}[h] 
\begin{center}
\includegraphics[width=.8\columnwidth]{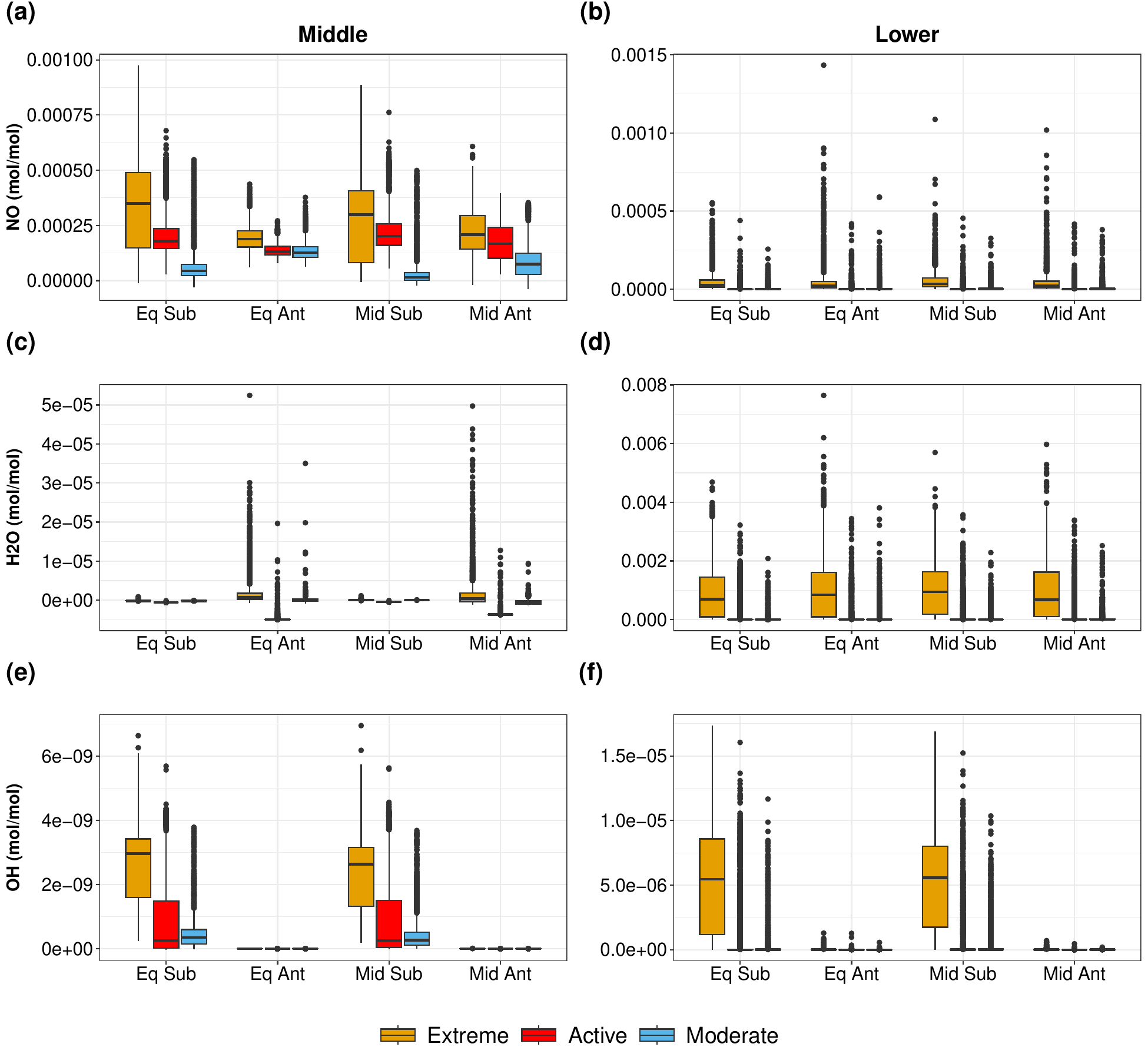}
\caption{\label{figS5} Same as Figure 2 but for Wind Speed (WS, m s$^{-1}$), Water Vapour (H$_2$O, mol mol$^{-1}$) and Nitrous Oxide (NO$_2$, mol mol$^{-1}$). Star indicates flare and large flare boxplots not significantly different at the 1\% level. } 
\end{center}
\end{figure*}

\begin{figure*}[h] 
\begin{center}
\includegraphics[width=.8\columnwidth]{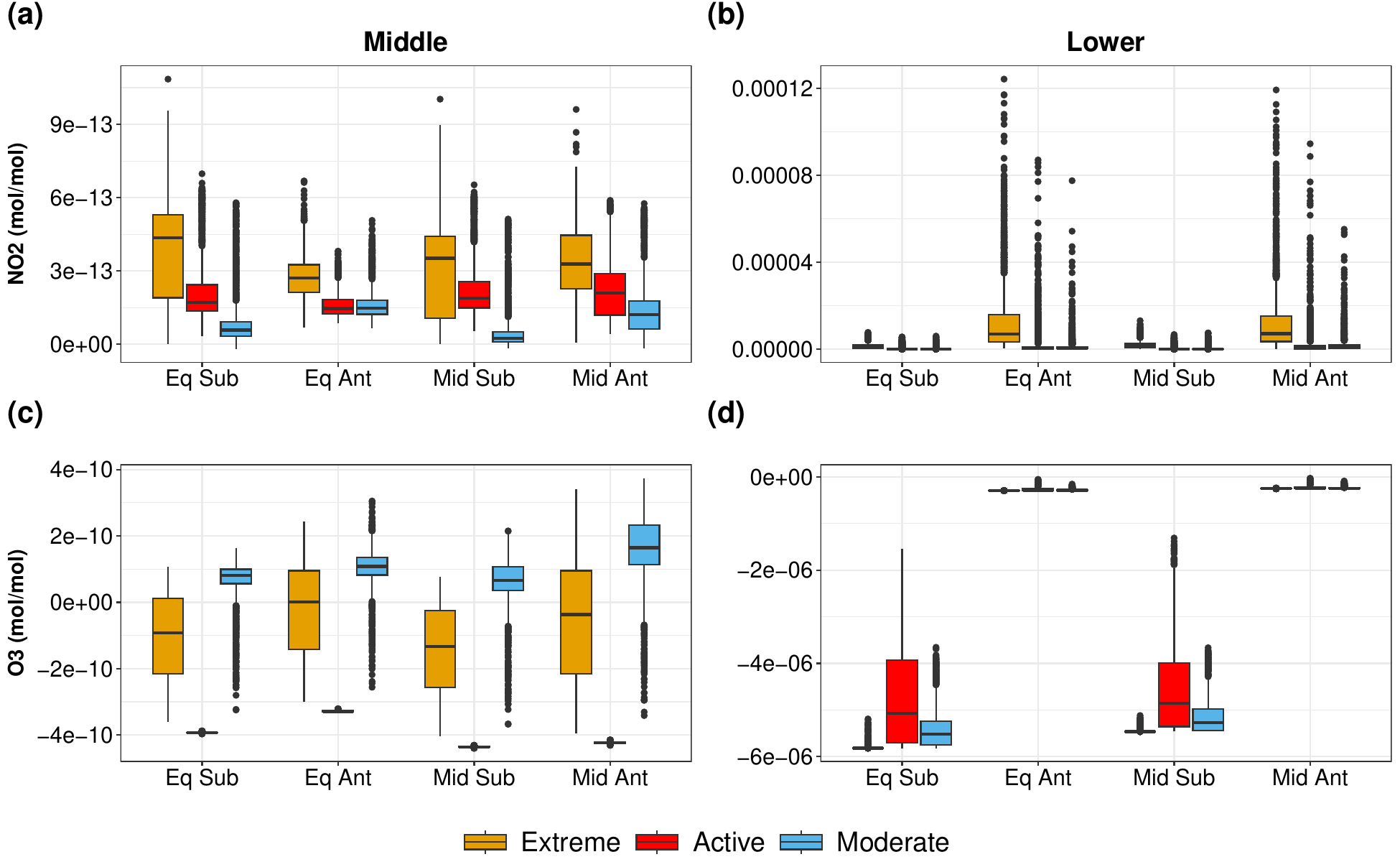}
\caption{\label{figS6}  Same as Figure 2, but for Nitrogen Dioxide (NO$_2$, mol mol$^{-1}$) and Ozone (O$_3$, mol mol$^{-1}$).} 
\end{center}
\end{figure*}

\begin{figure*}[h] 
\begin{center}
\includegraphics[width=.8\columnwidth]{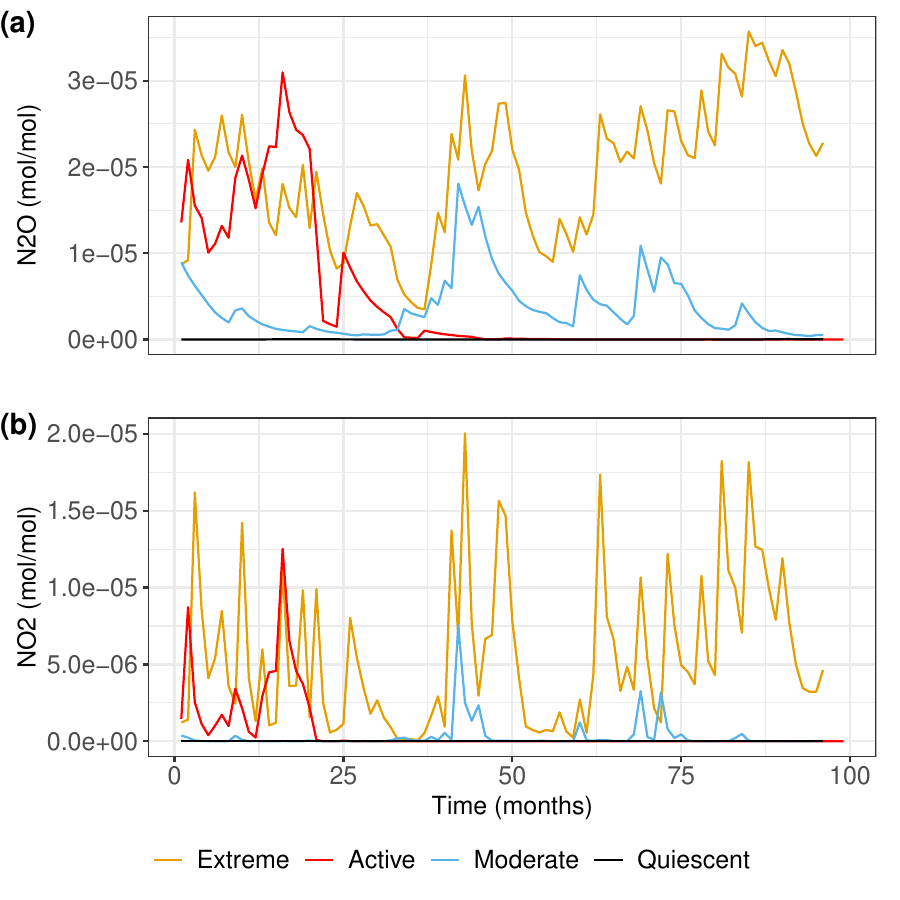}
\caption{\label{figS7}  Same as Figure 3 but for N$_2$O (a) and NO$_2$ (b).} 
\end{center}
\end{figure*}

\end{document}